\documentclass[oneside,a4paper,reqno]{amsart}
\usepackage{amssymb}
\newcommand{\myStrut}{\parbox{0.11 pt}{\rule{0 ex}{7.5 ex}}}

\newtheorem{lemma}{Lemma}
\newtheorem{theorem}{Theorem}
\theoremstyle{remark}
\newtheorem*{remark}{Remark}
\DeclareMathOperator{\Tr}{Tr}

\DeclareMathOperator{\Det}{Det}
\DeclareMathOperator{\U}{U}
\DeclareMathOperator{\Uc}{U_c}
\DeclareMathOperator{\W}{GP}
\DeclareMathOperator{\C}{C}
\DeclareMathOperator{\EC}{EC}
\DeclareMathOperator{\AC}{C^{*}}
\DeclareMathOperator{\Cc}{I}
\DeclareMathOperator{\SL}{SL}
\DeclareMathOperator{\ESL}{ESL}
\begin{document}
\begin{titlepage}
\begin{center}
\bfseries  SIC-POVMs AND THE EXTENDED CLIFFORD GROUP
\end{center}
\vspace{1 cm}
\begin{center} D M APPLEBY
\end{center}
\begin{center} Department of Physics, Queen Mary
University of London,  Mile End Rd, London E1 4NS,
UK
 \end{center}
\vspace{0.5 cm}
\begin{center}
  (E-mail:  D.M.Appleby@qmul.ac.uk)
\end{center}
\vspace{0.75 cm}
\vspace{1.25 cm}
\begin{center}
\vspace{0.35 cm}
\parbox{12 cm }{ 
We describe the structure of the extended Clifford Group (defined to be the group
consisting of all operators, unitary and anti-unitary, which normalize the
generalized Pauli group (or Weyl-Heisenberg group as it is often called)).  We
also obtain a  number of results concerning the structure of the Clifford Group
proper (\emph{i.e.}~the group consisting just of the unitary operators which
normalize the generalized Pauli group).  We then investigate the action of the
extended Clifford group operators on symmetric informationally complete POVMs (or
SIC-POVMs)  covariant relative to the action of the generalized Pauli group.  We
show that each of the fiducial vectors which has been constructed so far
(including all the vectors constructed numerically by Renes \emph{et al}) is an
eigenvector of one of a special class of order $3$ Clifford unitaries.  This
suggests a strengthening of a conjecture of Zauner's.  We  give a complete
characterization of the orbits and stability groups in dimensions $2$--$7$.
Finally, we  show that the problem of constructing fiducial vectors may be
expected to simplify in the infinite sequence of dimensions
$7, 13, 19, 21, 31,
\dots$.  We illustrate this point by constructing exact expressions for fiducial
vectors in dimensions $7$ and $19$.  
   }
\end{center}
\end{titlepage}
\section{Introduction}
\label{sec:Introduction}
The statistics of an arbitrary  quantum measurement are  described by
a positive operator valued measure, or POVM (Davies~\cite{Davies}, Busch \emph{et
al}~\cite{Busch1}, Peres~\cite{Peres}, Nielsen and Chuang~\cite{Nielsen} and
references cited therein). Suppose the measurement has only a finite number of
distinct outcomes.  Then the corresponding POVM assigns to each outcome $i$ 
the positive operator 
$\hat{E}_{i}$ with the property that
$\Tr (\hat{E}_i
\hat{\rho})$ is the probability of obtaining  outcome $i$
(where $\hat{\rho}$ is the density operator).  Since $\sum_{i} \Tr (\hat{E}_i
\hat{\rho})=1$ for all $\hat{\rho}$ we must have $\sum_{i} \hat{E}_i=1$.

A POVM is said to be \emph{informationally complete} if the probabilities
 $\Tr (\hat{E}_i
\hat{\rho})$ uniquely determine the density operator $\hat{\rho}$.  The concept
of informational  completeness is originally due to 
Prugove\v{c}ki~\cite{Prugo} (also see Busch~\cite{Busch2}, Busch \emph{et
al}~\cite{Busch1}, d'Ariano \emph{et al}~\cite{dAriano}, Flammia \emph{et
al}~\cite{Flammia}, Finkelstein~\cite{Finkelstein}, and references cited
therein).  It has an obvious relevance to the problem of quantum state
determination.  It also plays an important role in
 Caves \emph{et al}'s~\cite{Caves1,Caves2,FuchsSasaki,Fuchs} Bayesian approach to
the interpretation of quantum mechanics, and  in Hardy's~\cite{Hardy1,Hardy2}
proposed axiomatization. 

Suppose the Hilbert space has finite dimension $d$.  Then it is easily seen that
an informationally complete POVM must contain at least $d^2$ distinct operators
$\hat{E}_i$.  An informationally complete POVM is said to be \emph{symmetric
informationally complete} (or SIC) if it contains exactly this minimal number of
distinct operators and if, in addition,
\begin{enumerate}
\item $\lambda \hat{E}_i$ is a
one dimensional projector for all $i$ and some fixed constant $\lambda$.
\item The overlap $\Tr(\hat{E}_i \hat{E}_j)$ is the same for every pair of
distinct labels $i, j$.
\end{enumerate}
It is straightforward to show that this is equivalent to the requirement that, for
each
$i$,
\begin{equation}
 \hat{E}_i = \frac{1}{d} \left| \psi_i \right> \left< \psi_i \right|
\end{equation}
where the $d^2$  vectors $|\psi_i\rangle$ satisfy
\begin{equation}
\left|\langle \psi_i | \psi_j \rangle \right| = \begin{cases}
1 \qquad & i=j\\
\frac{1}{\sqrt{d+1}} \qquad & i\ne j
\end{cases}
\label{eq:OverlapCondition}
\end{equation}
SIC-POVMs were introduced in a dissertation by Zauner~\cite{Zauner}, and in
Renes
\emph{et al}~\cite{Renes}.  Wootters~\cite{Wootters1}, 
Bengtsson and Ericsson~\cite{Bengtsson,BengtssonB}
and Grassl~\cite{Grassl} have made further contributions.   There appear to be
some intimate connections with the theory of mutually unbiased
bases~\cite{Wootters1,Wootters2,Wootters3}, finite affine
planes~\cite{Wootters1,Bengtsson,BengtssonB}, and
polytopes~\cite{Bengtsson,BengtssonB}.

If SIC-POVMs existed in every finite dimension (or, failing that, in a
sufficiently large set of finite dimensions) they would constitute  a naturally
distinguished class of POVMs which might be expected to have many interesting
applications to quantum tomography, cryptography and information theory
generally.  They would also be  obvious candidates for the ``fiducial'' or
``standard'' POVMs  featuring in the work of Fuchs~\cite{Fuchs} and
Hardy~\cite{Hardy1,Hardy2}.

The question consequently arises:  is it in fact true that SIC-POVMs exist in
every finite dimension?  The answer to this question is still unknown.   Analytic
solutions  to Eqs.~(\ref{eq:OverlapCondition})
have been constructed in  dimensions
$2,3,4,5,6$ and $8$.  Moreover Renes \emph{et al}~\cite{Renes} have constructed
numerical solutions in dimensions $5$ to $45$ (the actual vectors can be
downloaded from their website~\cite{RenesVectors}).  So one may plausibly
speculate that SIC-POVMs exist in every finite dimension.  But it has not been
proved.

The SIC-POVMs which have so far been 
explicitly\footnote{
  Renes \emph{et al}~\cite{Renes} mention that they have constructed numerical
solutions 
  which are covariant under the action of other groups, but they do not give any
details.
 }
described in the literature are 
all covariant under the action of the generalized Pauli group (or Weyl-Heisenberg
group, as it is often called).  It is therefore natural to investigate their
behaviour under the action of the extended Clifford group.  The Clifford group
proper is defined to be the normalizer of the generalized Pauli group, considered
as a subgroup of $\U(d)$ (the group consisting  of all unitary operators in
dimension $d$).  It is relevant to a number of areas of quantum
information theory, and it has been extensively discussed in the
literature~\cite{Gottesman1,Gottesman2,Gottesman3,Dehaene,Hostens,vanDenNest,vanDenNestB}. 
Its relevance to the SIC-POVM problem has been stressed by Grassl~\cite{Grassl}.
As Grassl notes, it is related to the Jacobi group~\cite{JacobiRef}, which has
attracted some notice in the pure mathematical literature.  We define the
extended Clifford group to   be the group which results when the Clifford group
is enlarged, so as to include all
\emph{anti}-unitary operators which normalize the generalized  Pauli group.  As
we will see, this enlargement is essential if one wants to achieve a full
understanding of the SIC-POVM problem.

In Sections~\ref{sec:GPGroup}--\ref{sec:ExtendedClifford} we give a
self-contained account of  the structure of the extended Clifford group.  In the
course of this discussion we obtain a number of results concerning the  structure
of the Clifford group proper which, to the best of our knowledge, have not
previously appeared in the literature and which may be of some independent
interest.

In Section~\ref{sec:CliffordTrace} we define and establish some of the properties
of a function we call the Clifford trace.  We also identify a distinguished class
of order $3$ Clifford unitaries for which the Clifford trace $=-1$.  We refer to
these as \emph{canonical} order $3$ unitaries.  

In Section~\ref{sec:RBSCVectors} we analyze the vectors constructed numerically
by Renes \emph{et al}~\cite{Renes} (RBSC in the sequel) in dimension $5$--$45$. 
We show that each of them is an eigenvector of a canonical order $3$ Clifford
unitary.  This suggests the conjecture, that \emph{every} GP fiducial vector
is an eigenvector of a canonical order $3$ unitary.   We also show that, with
one exception, the stability group of each RBSC vector is order~$3$ (the
exception being dimension
$7$, where the stability group is order~$6$).

In Section~\ref{sec:Zauner} we show that RBSC's results also support  a
strengthened version of a conjecture of Zauner's~\cite{Zauner} (also see
Grassl~\cite{Grassl}).  

In Section~\ref{sec:VectorsOrbitsStability} we use RBSC's numerical data,
regarding the total number of fiducial vectors in dimensions $2$--$7$, to give a
complete characterization of the orbits and stability groups in dimensions
$2$-$7$.  Our results show that in each of these dimensions \emph{every} fiducial
vector covariant under the action of the generalized Pauli group is an
eigenvector of a canonical order $3$ Clifford unitary.  We also identify the
total number of distinct orbits.  It was already known~\cite{Renes,Grassl} that
there are infinitely many orbits in dimension~$3$, and one orbit  in
dimensions $2$ and $6$.  We show that there is, likewise, only one orbit in
dimensions $4$ and $5$, but two distinct orbits in dimension $7$.  We also
construct exact expressions for two fiducial vectors in dimension $7$ (one on
each of the two distinct orbits).

RBSC's numerical data may suggest that, after dimension $7$, the stability group
of every fiducial vector has order $3$.  In Section~\ref{sec:dimension19} we show
that there is at least one exception to that putative rule by constructing an
exact expression for a fiducial vector in dimension $19$ for which the stability
group has order $\ge 18$.

Our construction of  exact solutions in dimensions $7$ and $19$ was
facilitated by the fact that in these dimensions there exist canonical order $3$
unitaries having a particularly simple form.  In Section~\ref{sec:diagonalF} we
show that a similar simplification occurs in every dimension $d$ for which (a)
$d$ has at least one prime factor $= 1 \;(\text{mod}\; 3)$, (b) $d$ has no prime
factors $= 2 \;(\text{mod}\; 3)$ and (c) $d$ is not divisible by $9$.   In other
words, it happens when $d=7, 13, 19, 21, 31, \dots $.

\section{Fiducial Vectors for the Generalized Pauli Group}
\label{sec:GPGroup}
The SIC-POVMs which have been constructed to date all have a certain group
covariance property.  Let $G$ be a finite group having $d^2$ elements,
and suppose we have an injective map $g  \to \hat{U}_g $ which associates to each
 $g\in G$ a unitary operator $\hat{U}_g$ acting on $d$-dimensional Hilbert space.
Suppose that for all $g$, $g'$
\begin{equation}
 \hat{U}_{g} \hat{U}_{g'} = e^{i \xi_{g g'}} \hat{U}_{g g'}
\end{equation}
where $e^{i \xi_{g g'}}$ is a phase (so the map defines a group homomorphism of
$G$ into the quotient group $\U(d)/\Uc(d)$, where
$\Uc(d)$ is the centre of $\U(d)$).  Finally (and this, of course, is the
difficult part) suppose we  can find a  vector $|\psi\rangle\in
\mathbb{C}^d$ such that $\langle\psi|\psi\rangle=1$ and 
\begin{equation}
 \bigl| \langle \psi | \hat{U}_g | \psi\rangle \bigr| = \frac{1}{\sqrt{d+1}}
\end{equation}
for all $g\neq e$ ($e$ being the identity of $G$).  Then 
the assignment
\begin{equation}
 \hat{E}_{g} = \frac{1}{d} \hat{U}_{g} |\psi\rangle \langle \psi|
\hat{U}_{g}^{\dagger}
\end{equation}
defines a SIC-POVM on $\mathbb{C}^{d}$.  The vector $|\psi\rangle$ is said to be a
\emph{fiducial vector}. 

To date attention has been largely focussed on the case
$G=(\mathbb{Z}_d)^2$, where $\mathbb{Z}_d$ is the set of integers
$0,1,\dots, d-1$ under addition \emph{modulo} $d$ (although there is numerical
evidence that fiducial vectors exist for other choices of group~\cite{Renes}). 
That is also the case on which we will focus here.

To construct a suitable map  $(\mathbb{Z}_d)^2 \to \U(d)$, let $|e_0\rangle,
|e_1\rangle, \dots |e_{d-1}\rangle$ be an orthonormal basis for $\mathbb{C}^d$,
and let $\hat{T}$ be the  operator defined by
\begin{equation}
 \hat{T} |e_r\rangle =\omega^r |e_r\rangle
\label{eq:TDef}
\end{equation} 
 where $\omega = e^{2 \pi i/d}$.  Let $\hat{S}$ be the  shift operator
\begin{equation}
  \hat{S} |e_r\rangle = \begin{cases} 
  |e_{r+1}\rangle \qquad & r=0,1,\dots , d-2
\\
  |e_0\rangle  \qquad & r=d-1
\end{cases}
\label{eq:SDef}
\end{equation}
Then define, for each pair of integers $\mathbf{p}=(p_1,p_2)\in \mathbb{Z}^2$,
\begin{equation}
  \hat{D}_{\mathbf{p}} =\tau^{p_1 p_2} \hat{S}^{p_1} \hat{T}^{p_2}
\label{eq:DOpDefinition}
\end{equation}
where $\tau = - e^{\pi i /d}$ (the minus sign means  that
$\tau^{d^2}=1$ for all
$d$, thereby simplifying some of the formulae needed in the sequel).  We have,
for all $\mathbf{p}, \mathbf{q} \in \mathbb{Z}^2 $,
\begin{align}
 \hat{D}_{\mathbf{p}}^\dagger & = \hat{D}_{-\mathbf{p}} 
\label{eq:DConjugate}
\\
  \hat{D}_{\mathbf{p}} \hat{D}_{\mathbf{q}} & = \tau^{\langle
\mathbf{p},\mathbf{q}\rangle} \hat{D}_{\mathbf{p} + \mathbf{q}}
\label{eq:DCompositionRule}
\intertext{and}
\hat{D}_{\mathbf{p} + d\mathbf{q}} & = \begin{cases}
\hat{D}_{\mathbf{p}} \qquad & \text{if $d$ is odd}
\\
(-1)^{\langle \mathbf{p},\mathbf{q}\rangle} \hat{D}_{\mathbf{p}}
\qquad & \text{if $d$ is even}
\end{cases}
\label{eq:DShiftExpression}
\end{align}
where $\langle
 \mathbf{p},\mathbf{q}\rangle$ is the symplectic form
\begin{equation}
\langle
 \mathbf{p},\mathbf{q}\rangle=p_2 q_1-p_1 q_2
\end{equation}
  Consequently the map $\mathbf{p}
\in (\mathbb{Z}_d)^2  \to \hat{D}_{\mathbf{p}} \in \U(d)$
 has
all the required properties.  The operators $\hat{D}_{\mathbf{p}}$ are sometimes
called generalized Pauli matrices.  So we will say that a  vector
$|\psi\rangle\in \mathbb{C}^d$ is a generalized Pauli fiducial vector, or
\emph{GP fiducial vector} for short, if it is a fiducial vector relative to the
action of these operators:  \emph{i.e.}~if $\langle \psi |\psi \rangle =1$ and
\begin{equation}
\bigl| \langle \psi | \hat{D}_{\mathbf{p}} |\psi \rangle
\bigr| = \frac{1}{\sqrt{d+1}}
\label{eq:GPfiducial}
\end{equation}
for every $\mathbf{p} \in \mathbb{Z}^2  \neq \boldsymbol{0}\;(\text{mod}\;d)$.

The set of operators $\hat{D}_{\mathbf{p}}$ is not a group.  However, it becomes
a group if we allow each $\hat{D}_{\mathbf{p}}$ to be multiplied by an arbitrary
phase.  We will refer to the group
 $\W(d) = \{e^{i \xi} \hat{D}_{\mathbf{p}}\colon \xi \in \mathbb{R},
\mathbf{p} \in
\mathbb{Z}^2\}$ so obtained as the generalized Pauli 
group\footnote{
  Also known as the Weyl-Heisenberg group.
 Our definition is, perhaps, slightly unconventional.  It would be more usual
  to define $\W(d) =\{\tau^{n}
 \hat{D}_{\mathbf{p}}\colon n
 \in
 \mathbb{Z},
 \mathbf{p} \in
 \mathbb{Z}^2\}$---\emph{i.e.}\ the subgroup generated by the operators
  $\hat{D}_{\mathbf{p}}$.
}. 

We now want to investigate the normalizer of $\W(d)$:  \emph{i.e.}\ the
group $\C(d)$ consisting of all unitary operators $\hat{U}\in \U(d)$ with
the property
\begin{equation}
  \hat{U} \W(d) \hat{U}^{\dagger} = \W(d)
\end{equation}
The significance of this group for us is that it generates automorphisms of
$\W(d)$ according to the prescription
\begin{equation}
 \hat{P} \to \hat{U} \hat{P} \hat{U}^{\dagger}
\end{equation}
Consequently, if $|\psi\rangle$ is a GP fiducial vector, then so is $\hat{U}
|\psi\rangle$ for every
$\hat{U}\in
\C(d)$.

The group $\C(d)$ is known as the Clifford group, and has been
extensively discussed in the
literature~\cite{Gottesman1,Gottesman2,Gottesman3,Dehaene,Hostens,vanDenNest,vanDenNestB}.
Its relevance to the SIC-POVM problem has been stressed by 
 Grassl~\cite{Grassl}.  However, none of these accounts
derive  all the results  needed for our analysis of the RBSC
 vectors.  In the interests of
readability we give a unified treatment in the next section.
\section{The Clifford Group:  Structure, and Calculation of the Unitaries}
\label{sec:CliffordGroup}
  We begin with some definitions.  Let
\begin{equation}
  \overline{d}=\begin{cases}  d \qquad & \text{if $d$ is odd}
\\
2 d  \qquad & \text{if $d$ is even}
\end{cases}
\end{equation}
Let $\SL(2, \mathbb{Z}_{\overline{d}})$ be the group consisting of all $2\times 2$
matrices
\begin{equation}
\begin{pmatrix}
\alpha & \beta \\ \gamma & \delta
\end{pmatrix}
\end{equation}
such that $\alpha, \beta, \gamma, \delta \in \mathbb{Z}_{\overline{d}}$ and 
$\alpha \delta - \beta \gamma = 1 \; (\text{mod} \; \overline{d})$.  Note
that inverses exist in this group because the condition
$\alpha \delta - \beta \gamma = 1 \; (\text{mod} \; \overline{d})$ implies
\begin{equation}
\begin{pmatrix}
\alpha &  \beta \\ \gamma & \delta
\end{pmatrix}
\begin{pmatrix}
\delta & -\beta \\ -\gamma & \alpha
\end{pmatrix}
=
\begin{pmatrix}
1 & 0\\ 0 & 1
\end{pmatrix}
\end{equation}
in arithmetic \emph{modulo} $\overline{d}$.

We then have
\begin{lemma}
\label{lem:CliffordStructure1}
For each unitary operator $\hat{U}\in\C(d)$ there exists a  matrix $F\in
\SL(2,\mathbb{Z}_{\overline{d}})$  and  a vector
$\boldsymbol{\chi}\in (\mathbb{Z}_d)^2$ such that
\begin{equation}
 \hat{U} \hat{D}_{\mathbf{p}} \hat{U}^{\dagger}
= \omega^{\langle \boldsymbol{\chi}, F\mathbf{p}\rangle} \hat{D}_{F\mathbf{p}}
\end{equation}
for all $\mathbf{p}\in\mathbb{Z}^2$ (where $\omega = \tau^2=e^{2 \pi i/d}$, as
before).
\label{lem:FChiToAisSurjective}
\end{lemma}
\begin{proof}
If $\hat{U}\in\C(d)$ it is immediate that there exist functions $f$ and
$g$ such that 
\begin{equation}
 \hat{U} \hat{D}_{\mathbf{p}} \hat{U}^{\dagger}
=e^{ i g(\mathbf{p})} \hat{D}_{f(\mathbf{p})}
\label{eq:Lem1EqA}
\end{equation}
for all $\mathbf{p}\in \mathbb{Z}^2$.  It follows from
Eq.~(\ref{eq:DCompositionRule}) that
\begin{equation}
\left( e^{ i g(\mathbf{p})} \hat{D}_{f(\mathbf{p})} \right)
\left( e^{ i g(\mathbf{q})} \hat{D}_{f(\mathbf{q})} \right)
=\tau^{\langle{\mathbf{p},\mathbf{q}\rangle}}
\left(e^{ i g(\mathbf{p}+\mathbf{q})} \hat{D}_{f(\mathbf{p}+\mathbf{q})}
\right)
\end{equation}
for all $\mathbf{p}$, $\mathbf{q}\in\mathbb{Z}^2$.  Consequently
\begin{equation}
 e^{ i ( g(\mathbf{p})+g(\mathbf{q}))} \tau^{\langle
f(\mathbf{p}),f(\mathbf{q})\rangle} \hat{D}_{f(\mathbf{p})+f(\mathbf{q})}
= e^{ i g(\mathbf{p}+\mathbf{q})} \tau^{\langle{\mathbf{p},\mathbf{q}\rangle}}
\hat{D}_{f(\mathbf{p}+\mathbf{q})}
\label{eq:Lem1EqB}
\end{equation}
which implies $f(\mathbf{p}+\mathbf{q}) = f(\mathbf{p})+f(\mathbf{q})
\;(\text{mod}\; d)$.  We may therefore write
\begin{equation}
f(\mathbf{p}) = F' \mathbf{p} + d h(\mathbf{p})
\end{equation}
for some matrix $F'$ and function $h$.  Inserting this expression in
Eq.~(\ref{eq:Lem1EqA}) gives, in view of Eq.~(\ref{eq:DShiftExpression}),
\begin{equation}
\hat{U} \hat{D}_{\mathbf{p}} \hat{U}^{\dagger}
= e^{ i g(\mathbf{p})} \hat{D}_{F'\mathbf{p}+ d h(\mathbf{p})}
=\begin{cases}
e^{ i g(\mathbf{p})} \hat{D}_{F'\mathbf{p}} \qquad & \text{$d$ odd}
\\
e^{ i g(\mathbf{p})} (-1)^{\langle\mathbf{p},h(\mathbf{p})\rangle}
\hat{D}_{F'\mathbf{p}} \qquad & \text{$d$ even}
\end{cases}
\end{equation}
With the appropriate definition of $g'$ this means
\begin{equation}
 \hat{U} \hat{D}_{\mathbf{p}} \hat{U}^{\dagger}
=e^{ i g'(\mathbf{p})} \hat{D}_{F'\mathbf{p}}
\label{eq:Lem1EqC}
\end{equation}
for all $\mathbf{p}$.  Repeating the argument which  led to
Eq.~(\ref{eq:Lem1EqB}) we find
\begin{equation}
  e^{i g'(\mathbf{p}+\mathbf{q})-g'(\mathbf{p})-g'(\mathbf{q})}
\tau^{\langle \mathbf{p},\mathbf{q}\rangle - \langle
F'\mathbf{p},F'\mathbf{q}\rangle}=1
\label{eq:Lem1EqD}
\end{equation}
Interchanging $\mathbf{p}$ and $\mathbf{q}$ gives
\begin{equation}
  e^{i g'(\mathbf{p}+\mathbf{q})-g'(\mathbf{p})-g'(\mathbf{q})}
\tau^{-\langle \mathbf{p},\mathbf{q}\rangle + \langle
F'\mathbf{p},F'\mathbf{q}\rangle}=1
\end{equation}
We consequently require
\begin{equation}
\omega^{\langle \mathbf{p},\mathbf{q}\rangle - \langle
F'\mathbf{p},F'\mathbf{q}\rangle}
= \tau^{2\left( \langle \mathbf{p},\mathbf{q}\rangle - \langle
F'\mathbf{p},F'\mathbf{q}\rangle \right)}
=1
\end{equation}
for all $\mathbf{p}, \mathbf{q}$.  It is readily verified that 
$\langle
F'\mathbf{p},F'\mathbf{q}\rangle =(\Det F') \langle
\mathbf{p},\mathbf{q}\rangle $.  We must therefore have
\begin{equation}
  \Det F' = 1 \; (\text{mod} \; d)
\end{equation}
If $d$ is odd, or if $d$ is even and $\Det F' = 1\; (\text{mod} \;
\overline{d}) $, we can find a matrix $F\in \SL(2,\mathbb{Z}_{\overline{d}})$ such
that 
$F=F' \; (\text{mod} \;
\overline{d})$.  It then follows from Eq.~(\ref{eq:DShiftExpression}) that
$\hat{D}_{F\mathbf{p}}=\hat{D}_{F'\mathbf{p}}$ for all $\mathbf{p}$.

Suppose, on the other hand,
$d$ is even and 
$\Det F' \neq 1\; (\text{mod} \; \overline{d}) $.  Then
$\Det F' = d+1 \; (\text{mod} \; \overline{d})$.  Write
\begin{equation}
  F'=
\begin{pmatrix}
\alpha & \beta \\ \gamma & \delta
\end{pmatrix}
\end{equation}
We know $\alpha \delta -\beta \gamma = \Det F'$ is odd.  So either $\alpha,
 \delta$ are both odd, or else $\beta, \gamma$ are both odd.  If $\alpha, \delta$
are both odd let
\begin{equation}
  \Delta=
\begin{pmatrix}
1 & 0 \\ 0 & 0
\end{pmatrix}
\end{equation}
while if $\beta, \gamma$ are both odd let
\begin{equation}
  \Delta=
\begin{pmatrix}
0 & 1 \\ 0 & 0
\end{pmatrix}
\end{equation}
Then $\Det (F'+d \Delta) = 1\;(\text{mod}\; \overline{d})$.  We can therefore
choose a matrix $F\in \SL(2,\mathbb{Z}_{\overline{d}})$ such
that 
$F=F'+d \Delta \; (\text{mod} \;
\overline{d})$.  Inserting this expression in Eq.~(\ref{eq:Lem1EqC}) we have, in
view of Eq.~(\ref{eq:DShiftExpression}),
\begin{equation}
 \hat{U} \hat{D}_{\mathbf{p}} \hat{U}^{\dagger}
=e^{ i g'(\mathbf{p})} \hat{D}_{(F-d \Delta )\mathbf{p}} = e^{ i g'(\mathbf{p})}
(-1)^{\langle F\mathbf{p},\Delta \mathbf{p}\rangle} \hat{D}_{F\mathbf{p}}
\end{equation}
We conclude that there is, in every case, a function $g''$ and a matrix 
$F\in \SL(2,\mathbb{Z}_{\overline{d}})$ such that
\begin{equation}
\hat{U} \hat{D}_{\mathbf{p}} \hat{U}^{\dagger}
=
e^{ i g''(\mathbf{p})} \hat{D}_{F\mathbf{p}}
\end{equation}
for all $\mathbf{p}$. 

It remains to establish the form of the function $g''$.  We  note, first of all,
that it follows from Eqs.~(\ref{eq:DOpDefinition})
and~(\ref{eq:DCompositionRule}) that
\begin{equation}
\bigl( \hat{D}_{\mathbf{p}}\bigr)^d=\hat{D}_{d \mathbf{p}} = \tau^{d^2 p_1 p_2}
\hat{S}^{d p_1}
\hat{T}^{d p_2} =1
\label{eq:DpPower}
\end{equation}
for all $\mathbf{p}$ (because $\hat{S}^d=\hat{T}^d = \tau^{d^2} =1$).
Consequently
\begin{equation}
1  = \hat{U} \bigl( \hat{D}_{\mathbf{p}}\bigr)^d \hat{U}^{\dagger}
 =\bigl(\hat{U} \hat{D}_{\mathbf{p}} \hat{U}^{\dagger}\bigr)^d
=e^{ i d g''(\mathbf{p})} \bigl( \hat{D}_{F \mathbf{p}}\bigr)^d
= e^{ i d g''(\mathbf{p})}
\end{equation}
for all $\mathbf{p}$.  We must therefore have $e^{ i 
g''(\mathbf{p})}=\omega^{\tilde{g}(\mathbf{p})}$ for some function $\tilde{g}$
taking values in $\mathbb{Z}_d$. Repeating the argument which led to
Eq.~(\ref{eq:Lem1EqD}) we find
\begin{equation}
 \omega^{\tilde{g}(\mathbf{p}+\mathbf{q})-\tilde{g}(\mathbf{p})-\tilde{g}
(\mathbf{q})}
\tau^{\langle \mathbf{p},\mathbf{q}\rangle - \langle
F\mathbf{p},F\mathbf{q}\rangle}=1
\end{equation}
We have $\langle \mathbf{p},\mathbf{q}\rangle - \langle
F\mathbf{p},F\mathbf{q}\rangle = (1-\Det F)\langle \mathbf{p},\mathbf{q}\rangle
=0\; (\text{mod} \; \overline{d})$.  Consequently 
$\tau^{\langle \mathbf{p},\mathbf{q}\rangle - \langle
F\mathbf{p},F\mathbf{q}\rangle}=1$ (because $\tau^{\overline{d}}=1$) and so
\begin{equation}
  \tilde{g}(\mathbf{p}+\mathbf{q})=\tilde{g}(\mathbf{p})+\tilde{g}
(\mathbf{q}) \; (\text{mod}\; d)
\end{equation}
for all $\mathbf{p}, \mathbf{q}$.
This implies $\tilde{g}(\mathbf{p})=\langle \boldsymbol{\chi}',\mathbf{p}\rangle
\; (\text{mod}\; d)$ for for all $\mathbf{p}$ some fixed
$\boldsymbol{\chi}' \in (\mathbb{Z}_d)^2$.  Setting
$\boldsymbol{\chi}=F \boldsymbol{\chi}'$, and using the fact that $\langle
F^{-1}\boldsymbol{\chi},\mathbf{p}\rangle=\langle
\boldsymbol{\chi},F\mathbf{p}\rangle\; (\text{mod}\; d)$ we conclude
\begin{equation}
\hat{U} \hat{D}_{\mathbf{p}} \hat{U}^{\dagger}
=
\omega^{\langle
\boldsymbol{\chi},F\mathbf{p}\rangle} \hat{D}_{F\mathbf{p}}
\end{equation}
for all $\mathbf{p}$.
\end{proof}
We now want to prove the converse of Lemma~\ref{lem:FChiToAisSurjective}.  That
is, we want to prove that, for each pair $F\in\SL(2,\mathbb{Z}_{\overline{d}})$
and
$\boldsymbol{\chi}\in (\mathbb{Z}_d)^2$ there is a corresponding operator
$\hat{U}\in\C(d)$.  We also want to derive an explicit expression for the
operator $\hat{U}$ (this has, in effect, already been done by 
Hostens~\emph{et
al}~\cite{Hostens}; however, the formulae we derive are different, and better
adapted to the questions addressed in this paper).

We begin by focussing on a special class of matrices $F$.  Let
$[n_1,n_2,\dots,n_r]$ denote the GCD (greatest common divisor) of the integers
$n_1,n_2, \dots, n_r$.  We define the class of \emph{prime matrices} to be the
set of all matrices
\begin{equation}
  F=
\begin{pmatrix}
  \alpha & \beta \\ \gamma & \delta
\end{pmatrix}
\end{equation}
$\in \SL(2, \mathbb{Z}_{\overline{d}})$ such that $\beta$ is non-zero and 
$[\beta,
\overline{d}]=1$ (so that $\beta$ has a multiplicative inverse in
$\mathbb{Z}_{\overline{d}}$).  We then have
\begin{lemma}
\label{lem:CliffordStructure2}
Let
\begin{equation}
  F=
\begin{pmatrix}
  \alpha & \beta \\ \gamma & \delta
\end{pmatrix}
\end{equation}
be a prime matrix $\in \SL(2, \mathbb{Z}_{\overline{d}})$.    Let
\begin{equation}
\hat{V}_{F} = \frac{1}{\sqrt{d}} \sum_{r,s=0}^{d-1} 
\tau^{\beta^{-1} \left(\alpha s^2 - 2 r s+ \delta r^2  \right)} |e_r\rangle
\langle e_s|
\label{eq:VFDef}
\end{equation}
(where
$\beta^{-1}\in \mathbb{Z}_{\overline{d}}$ is such that $\beta^{-1}\beta = 1 \;
(\text{mod}\;
\overline{d})$).  Then $\hat{V}_F$ is a unitary operator $\in\C(d)$ such
that
\begin{equation}
 \hat{V}_{F}^{\vphantom{\dagger}} \hat{D}_{\mathbf{p}}^{\vphantom{\dagger}}
\hat{V}_{F}^{\dagger} = \hat{D}_{F\mathbf{p}}^{\vphantom{\dagger}}
\end{equation}
for all $\mathbf{p}$.
\end{lemma}
\begin{proof}
Let 
\begin{equation}
\hat{S}' = \hat{D}_{(\alpha,\gamma)} \qquad \text{and} \qquad
\hat{T}' = \hat{D}_{(\beta,\delta)} 
\end{equation}
and define
\begin{equation}
|f_{0}\rangle = \frac{1}{\sqrt{d}} \sum_{r=0}^{d-1} (\hat{T}')^r |e_0\rangle
\end{equation}
It follows from Eq.~(\ref{eq:DpPower}) that $\bigl( \hat{T}' \bigr)^{d} =
 1$.
Consequently
\begin{equation}
 \hat{T}'|f_{0}\rangle = |f_{0}\rangle
\end{equation}
It follows from Eq.~(\ref{eq:DCompositionRule}) that 
$\hat{T}' \hat{S}'=\omega \hat{S}' \hat{T}'$.  So we can obtain a complete set of
eigenvectors by laddering.  Specifically, let 
\begin{equation}
 |f_r\rangle = \bigl(\hat{S}'\bigr)^{r} | f_0\rangle
\end{equation}
for $r=1,\dots, d-1$.  Then
\begin{equation}
  \hat{T}' |f_r\rangle = \omega^{r} |f_r\rangle
\label{eq:Lem2EqB}
\end{equation}
for all $r$.  Since $\bigl( \hat{S}' \bigr)^{d} =
 1$ (as follows from Eq.~(\ref{eq:DpPower})) we also have
\begin{equation}
  \hat{S}' |f_r\rangle = | f_{r \oplus_d 1}\rangle
\end{equation}
for all $r$ (where $\oplus_d$ signifies addition \emph{modulo} $d$).

We next show that the vectors $|f_r\rangle$ are orthonormal.  It follows from
Eqs.~(\ref{eq:TDef}), (\ref{eq:SDef}), (\ref{eq:DOpDefinition})
and~(\ref{eq:DCompositionRule}) that
\begin{equation}
 (\hat{T}')^r |e_{0}\rangle
=
 \hat{D}_{(r\beta,r\delta)} |e_{0}\rangle
=
\tau^{ \beta \delta r^2} \hat{S}^{\beta r} |e_{0}\rangle
\end{equation}
and consequently
\begin{equation}
|f_0\rangle=\left(\frac{1}{\sqrt{d}} \sum_{r=0}^{d-1} \tau^{ \beta \delta r^2}
\hat{S}^{\beta r}\right) |e_{0}\rangle
=
\left(\frac{1}{\sqrt{d}} \sum_{r=0}^{d-1} \tau^{ \beta^{-1} \delta (\beta r)^2}
\hat{S}^{\beta r}\right) |e_{0}\rangle
\end{equation}
(where we have used the fact that $\tau^{\overline{d}}=1$).
We need to be  careful at this point, due to the fact that congruence
\emph{modulo} $d$ need not imply congruence \emph{modulo} $\overline{d}$.  Let
$q_r$ be the quotient of $\beta r$ on division by $d$, and let $t_r$ be the
remainder.   So $\beta r = q_r d + t_r$ and
\begin{equation}
|f_0\rangle=
\left(\frac{1}{\sqrt{d}} \sum_{r=0}^{d-1} \tau^{ \beta^{-1} \delta (q_r d +
t_r)^2}
\hat{S}^{q_r d + t_r}\right) |e_{0}\rangle
\end{equation}
We have 
\begin{align}
\hat{S}^{q_r d + t_r} & = \hat{S}^{t_r}
\intertext{and}
\tau^{ \beta^{-1} \delta (q_r d +
t_r)^2} & =
\tau^{\beta^{-1} \delta (t_r^2 + 2 d q_r t_r + d^2 t_r^2)}
=
\tau^{\beta^{-1} \delta t_r^2}
\end{align}
(because $\tau^{2 d}=\tau^{d^2} =1$).  Consequently
\begin{equation}
|f_0\rangle=
\left(\frac{1}{\sqrt{d}} \sum_{r=0}^{d-1} \tau^{ \beta^{-1} \delta 
t_r^2}
\hat{S}^{t_r}\right) |e_{0}\rangle
=
\frac{1}{\sqrt{d}}\sum_{r=0}^{d-1} \tau^{ \beta^{-1} \delta 
t_r^2}|e_{t_r}\rangle
\end{equation}
The fact that $[\beta,\overline{d}]=1$ implies that $[\beta,d]=1$. It follows
that, as
$r$ runs over the integers $0,1,\dots, d-1$, so does $t_r$ (though not
necessarily in the same order).  Consequently
\begin{equation}
|f_0\rangle=
\frac{1}{\sqrt{d}}\sum_{t=0}^{d-1} \tau^{ \beta^{-1} \delta 
t^2}|e_{t}\rangle
\label{eq:Lem2EqA}
\end{equation}
It follows that 
\begin{equation}
 \langle f_r|f_r\rangle = \langle f_0|( \hat{S}')^{-r} (\hat{S}')^{r} |
f_0\rangle = \langle f_{0} | f_{0}\rangle = 1
\end{equation}
The fact that $\langle f_r | f_s \rangle=0$ when $r\neq s$ is an immediate
consequence of the fact that $|f_r\rangle$, $|f_s \rangle$ are eigenvectors of
$\hat{T}'$ corresponding to different eigenvalues.  We conclude that
\begin{equation}
\langle f_r | f_s \rangle=\delta_{rs}
\end{equation}
as claimed.

We now want to calculate an explicit formula for $|f_r\rangle$ when $r>0$. It
follows from previous results that
\begin{equation}
|f_r\rangle 
= 
\hat{D}_{r\alpha,r\gamma} |f_0\rangle
=
\frac{1}{\sqrt{d}}\sum_{t=0}^{d-1} \tau^{ \beta^{-1} \delta 
t^2+\alpha \gamma r^2 + 2 \gamma r t} (\hat{S})^{r \alpha} |e_{t}\rangle
\end{equation}
By an argument similar to the one leading to Eq.~(\ref{eq:Lem2EqA}) we deduce
\begin{align}
|f_r\rangle 
& =
\frac{1}{\sqrt{d}}\sum_{t=0}^{d-1} \tau^{ \beta^{-1} \delta 
(t-\alpha r)^2+\alpha \gamma r^2 + 2 \gamma r (t-\alpha r)} 
|e_{t}\rangle
\\
& =
\frac{1}{\sqrt{d}}\sum_{t=0}^{d-1} \tau^{ \beta^{-1} \bigl( \delta t^2 -
2 r t + \alpha r^2 \bigr)}
|e_{t}\rangle
\end{align}
(since $\alpha \delta - \beta \gamma = 1 \;
(\text{mod}\; \overline{d})$).  Comparing with Eq.~(\ref{eq:VFDef}) we see that
\begin{equation}
 \hat{V}_{F} = \sum_{r=0}^{d-1} |f_r\rangle \langle e_r |
\end{equation}
which shows that $\hat{V}_{F}$ is unitary.  Moreover,
\begin{equation}
 \hat{V}_{F} \hat{T} \hat{V}_{F}^{\dagger} |f_r\rangle
=
\hat{V}_{F} \hat{T} |e_r\rangle
=\omega^{r} |f_r\rangle
\end{equation}
for all $r$.  Comparing with Eq.~(\ref{eq:Lem2EqB}) we deduce $\hat{V}_{F}
\hat{T}
\hat{V}_{F}^{\dagger}=\hat{T}'$.  Similarly
$\hat{V}_{F}
\hat{S}
\hat{V}_{F}^{\dagger}=\hat{S}'$.  Hence
\begin{align}
\hat{V}_{F}
\hat{D}_{\mathbf{p}}
\hat{V}_{F}^{\dagger}
& =
\tau^{p_1 p_2} \hat{V}_{F}
\hat{S}^{p_1} \hat{T}^{p_2}
\hat{V}_{F}^{\dagger}
\\
& =
\tau^{p_1 p_2} \hat{D}_{\alpha p_1,\gamma p_1} \hat{D}_{\beta p_2,\delta p_2}
\\
& = \tau^{\bigl(1-\beta\gamma+\alpha\delta  \bigr)p_1 p_2}
\hat{D}_{F \mathbf{p}}
\\
& =
\hat{D}_{F \mathbf{p}}
\end{align}
for all $\mathbf{p}$. 
\end{proof}
To extend this result to the case of an arbitary matrix $\in
\SL(2,\mathbb{Z}_{\overline{d}})$ we need the following decomposition lemma,
which states that every non-prime matrix can be written as the  product of two
prime matrices:
\begin{lemma}
\label{lem:CliffordStructure3}
Let 
\begin{equation}
  F=
\begin{pmatrix}
  \alpha & \beta \\ \gamma & \delta
\end{pmatrix}
\end{equation}
be a non-prime matrix $\in \SL(2, \mathbb{Z}_{\overline{d}})$.  Then there exists
an integer $x$ such that $\delta + x\beta$ is non-zero and $[\delta +
x\beta,\overline{d}]=1$.  Let $x$ be any integer having that property, and let
\begin{align}
F_1 & = \begin{pmatrix}
  0 & -1 \\ 1 & x
\end{pmatrix}
\\
F_2 & =  \begin{pmatrix}
  \gamma + x \alpha & \delta + x \beta \\  - \alpha & -\beta
\end{pmatrix}
\end{align}
Then $F_1$, $F_2$ are prime matrices $\in \SL(2, \mathbb{Z}_{\overline{d}})$ such
that
\begin{equation}
F=F_1 F_2
\end{equation}
\end{lemma}
\begin{proof}
Suppose, to begin with, that $\beta, \delta$ are both non-zero.  Let
$k=[\beta,\delta]$.  We then have
\begin{align}
\beta & = k \beta_0\\
\delta & = k \delta_0
\end{align}
where 
$[\beta_0,\delta_0]=1$.  We also have $[k,\overline{d}]=1$
(because $\alpha \delta-\beta \gamma = 1\; (\text{mod}\; \overline{d})$).  The
fact that $\beta_0$, $\delta_0$ are relatively prime means we can use Dirichlet's
theorem  (see, for example, Nathanson~\cite{Nathanson} or Rose~\cite{Rose}) to
deduce that the sequence
\begin{equation}
\delta_0, \; (\delta_0 +\beta_0),\; (\delta_0
+ 2 \beta_0), \dots
\end{equation}
contains infinitely many primes.  Consequently, there exists an integer $x$ such
that $\delta_0 + x \beta_0\neq0$ and $[\delta_0 + x
\beta_0, \overline{d}]=1$.  The fact that $k\neq 0$ and
$[k,\overline{d}]=1$ then implies that $\delta+ x\beta\neq 0$ and $[\delta +
x\beta,
\overline{d}]=1$.  The claim is now immediate.

It remains to consider the case when $\beta, \delta$ are not both  non-zero. 
If $\delta =0$ the fact that $\det F = 1 \; (\text{mod}\;\overline{d})$ would
imply that
$\beta
\neq 0$ and
$[\beta,
\overline{d}]=1$---contrary to the assumption that the matrix $F$ is non-prime. 
Suppose, on the other hand, that $\beta =0$.  Then the fact that
$\det F = 1 \; (\text{mod}\;\overline{d})$ implies that $\delta\neq 0$ and 
$[\delta,
\overline{d}]=1$.  So the claim is true for every choice of $x$.
\end{proof}
We can now deduce the following converse of Lemma~\ref{lem:CliffordStructure1}:
\begin{lemma}
\label{lem:CliffordStructure4}
Let $(F,\boldsymbol{\chi})$ be any pair $\in
\SL(2,\mathbb{Z}_{\overline{d}}) \times (\mathbb{Z}_d)^2$.  If $F$ is a prime
matrix define
\begin{equation}
\hat{U} = \hat{D}_{\boldsymbol{\chi}} \hat{V}_F
\label{eq:CliffordSurjection1}
\end{equation}
(where $\hat{V}_F$ is the operator defined by Eq.~(\ref{eq:VFDef})). 
If $F$ is non-prime choose  two prime matrices $F_1, F_2$ such that $F=F_1 F_2$
(the existence of such matrices being guaranteed by
Lemma~\ref{lem:CliffordStructure3}), and define
\begin{equation}
\hat{U} = \hat{D}_{\boldsymbol{\chi}} \hat{V}_{F_1} \hat{V}_{F_2}
\label{eq:CliffordSurjection2}
\end{equation}
(where $\hat{V}_{F_1}, \hat{V}_{F_2}$ are the operators defined by
Eq.~(\ref{eq:VFDef})).  Then
\begin{equation}
 \hat{U} \hat{D}_{\mathbf{p}} \hat{U}^{\dagger}
= \omega^{\langle \boldsymbol{\chi}, F\mathbf{p}\rangle} \hat{D}_{F\mathbf{p}}
\label{eq:UforGeneralFandChi}
\end{equation}
for all $\mathbf{p}\in\mathbb{Z}^2$
\end{lemma}
\begin{proof}
The claim is an immediate consequence of Eqs.~(\ref{eq:DConjugate}),
(\ref{eq:DCompositionRule}) and Lemma~\ref{lem:CliffordStructure2}.
\end{proof}
If $\hat{U}$, $\hat{U}'$ differ by a phase, so that $\hat{U}'=e^{i \theta}
\hat{U}$, they have the same action on the generalized Pauli group:
\begin{equation}
\hat{U} \hat{D}_{\mathbf{p}}
\hat{U}^{\dagger}=\hat{U}'\hat{D}_{\mathbf{p}} \hat{U}'\mathstrut^{\dagger}
\end{equation}
for all $\mathbf{p}$.
So the object of real interest is not the Clifford group itself, but the group
$\C(d)/\Cc(d)$ which results when the phases are factored out.  Here $\Cc(d)$ is
the subgroup  consisting of all operators of the form $e^{i \theta}
\hat{I}$, where $\hat{I}$ is the identity operator and $\theta\in \mathbb{R}$.
The elements of
$\C(d)/\Cc(d)$ are often called  \emph{Clifford
operations}.

Let $\SL(2, \mathbb{Z}_{\overline{d}}) \ltimes
(\mathbb{Z}_d)^2$ be the semi-direct product of
$\SL(2,\mathbf{Z}_{\overline{d}})$ and $(\mathbb{Z}_d)^2$: 
\emph{i.e.}~the group which results when the set 
$\SL(2,
\mathbb{Z}_{\overline{d}})
\times (\mathbb{Z}_d)^2$ is equipped  with the composition rule
\begin{equation}
(F_1,\boldsymbol{\chi}_1) \circ (F_2,\boldsymbol{\chi}_2)
= (F_1 F_2, \boldsymbol{\chi}_1 + F_1 \boldsymbol{\chi}_2)
\label{eq:SemiDirectCompositionRule}
\end{equation}
Then we have the following structure theorem, which states that $\C(d)/\Cc(d)$ is
naturally isomorphic to  $\SL(2, \mathbb{Z}_{\overline{d}}) \ltimes
(\mathbb{Z}_d)^2$ when $d$ is odd, and naturally isomorphic to a quotient group
of  $\SL(2, \mathbb{Z}_{\overline{d}}) \ltimes
(\mathbb{Z}_d)^2$
when
$d$ is even:
\begin{theorem}
\label{thm:CliffordStructure}
There exists a unique surjective homomorphism
\begin{equation}
f  \colon \SL(2, \mathbb{Z}_{\overline{d}}) \ltimes
(\mathbb{Z}_d)^2 \to \C(d)/\Cc(d)
\end{equation}
with the property $\hat{U}
\hat{D}_{\mathbf{p}}
\hat{U}^{\dagger} =\omega^{<\boldsymbol{\chi},F
\mathbf{p}>}\hat{D}_{F\mathbf{p}}$  for each $\hat{U} \in
f(F,\boldsymbol{\chi})$ and all
$\mathbf{p}\in \mathbb{Z}^2$.

If $d$ is odd $f$ is an isomorphism.  If $d$ is even the kernel of $f$ is the
subgroup  $K_f\subseteq \SL(2, \mathbb{Z}_{\overline{d}}) \ltimes
(\mathbb{Z}_d)^2$ consisting of the $8$ elements of the form
\begin{equation}
\left(\begin{pmatrix}1+ r d & s d \\ t d & 1+r d
\end{pmatrix},
\begin{pmatrix} s d /2 \\ t d/2
\end{pmatrix}
\right)
\label{eq:KernelExpression}
\end{equation}
where $r,s,t = 0$ or $1$. 
\end{theorem}
\begin{proof}
An operator $\hat{U}\in \C(d)$ has the property 
\begin{equation}
\hat{U} \hat{D}_{\mathbf{p}} \hat{U}^{\dagger} = \hat{D}_{\mathbf{p}}
\end{equation}
for all $\mathbf{p}$
if and only if it is a multiple of the identity.  So it follows from results
already proved that there is exactly one surjective map
\begin{equation}
f  \colon \SL(2, \mathbb{Z}_{\overline{d}}) \ltimes
(\mathbb{Z}_d)^2 \to \C(d)/\Cc(d)
\end{equation}
such that $\hat{U}
\hat{D}_{\mathbf{p}}
\hat{U}^{\dagger} =\omega^{<\boldsymbol{\chi},F
\mathbf{p}>}\hat{D}_{F\mathbf{p}}$  for each $\hat{U} \in
f(F,\boldsymbol{\chi})$ and all
$\mathbf{p}\in \mathbb{Z}^2$.  The fact that $f$ is actually a homomorphism is
then an immediate consequence of the definitions.

Let $K_f$ be the kernel of $f$.   Then $(F,\boldsymbol{\chi})\in K_f$ if and
only if
\begin{equation}
\omega^{<\boldsymbol{\chi},F\mathbf{p}>} \hat{D}_{F \mathbf{p}} =
\hat{D}_{\mathbf{p}}
\label{eq:KernelDef}
\end{equation}
for all $\mathbf{p}$.  For that to be true we must have $F=
1\; (\text{mod}\; d)$.  If $d$ is odd this implies
$\hat{D}_{F
\mathbf{p}} = \hat{D}_{\mathbf{p}}$ for all $\mathbf{p}$.   
Eq.~(\ref{eq:KernelDef}) then becomes
$\omega^{<\boldsymbol{\chi},\mathbf{p}>}=1$ for all $\mathbf{p}$, implying
$\boldsymbol{\chi}=\begin{pmatrix} 0 \\ 0\end{pmatrix}$.  So
the kernel is trivial, and
$f$ is an isomorphism as claimed.

Suppose, on the other hand, that $d$ is even.  The condition $F=
1\; (\text{mod}\; d)$
 then implies that $F=1+ d\Delta$, where $\Delta$ is a matrix of the form
\begin{equation}
\Delta=\begin{pmatrix} r_1 & s \\ t & r_2
\end{pmatrix}
\end{equation}
with $r_1, r_2, s, t= 0$ or $1$.   Inserting this expression in
Eq.~(\ref{eq:KernelDef}) we find, in view of
Eqs.~(\ref{eq:DConjugate}--\ref{eq:DShiftExpression}), that
$(F,\boldsymbol{\chi})\in K_f$ if and only if
\begin{equation}
1=\omega^{<\boldsymbol{\chi},F\mathbf{p}>} \hat{D}_{F \mathbf{p}}
\hat{D}_{-\mathbf{p}}
= \omega^{<\boldsymbol{\chi},\mathbf{p}>}
\tau^{d <\mathbf{p},\Delta \mathbf{p}>}
\end{equation}
for all $\mathbf{p}$.  After re-arranging the condition becomes
\begin{equation}
\omega^{\chi_2 p_1-\chi_1 p_2} = (-1)^{(r_1-r_2)p_1 p_2 - t p_1^2 + s p_2^2}
=(-1)^{(r_1-r_2)p_1 p_2 + t p_1 - s p_2}
\end{equation}
for all $\mathbf{p}$.  This is true if and only if $r_1=r_2$, $\chi_1=sd/2$ and
$\chi_2=td/2$.
\end{proof}
We conclude with a result concerning the order of the group $\C(d)/\Cc(d)$
which will be needed later on.  Let $\nu(n,d)$ be the number of distinct  ordered
pairs
$(x,y)\in (\mathbb{Z}_d)^2$ such that $x y = n\; (\text{mod}\; d)$.  We then have
\begin{lemma}
\label{lem:CliffordStructure5}
The order of the group $\C(d)/\Cc(d)$ is
\begin{equation}
\bigl| \C(d)/\Cc(d)\bigr|
=d^2 \left( \sum_{n=0}^{d-1} \nu(n,d)\nu(n+1,d) \right)
\label{eq:CliffordOrderFormula1}
\end{equation}
If $d$ is a prime number this reduces to
\begin{equation}
\bigl| \C(d)/\Cc(d)\bigr|=d^3 (d^2-1)
\label{eq:CliffordOrderFormula2}
\end{equation}
\end{lemma}
\begin{proof}
We begin by showing that $\C(d)/\Cc (d)$ and $\SL(2, \mathbb{Z}_d) \ltimes
(\mathbb{Z}_d)^2$ have the same cardinality when considered as \emph{sets}.  This
is true for all
$d$, notwithstanding the fact that when $d$ is even $\C(d)/\Cc (d)$ and $\SL(2,
\mathbb{Z}_d)
\ltimes (\mathbb{Z}_d)^2$ are not naturally isomorphic as \emph{groups}.

The statement is immediate when $d$ is odd.  Suppose, on the other hand, that $d$
is even.  Let $g: \SL(2,\mathbb{Z}_{2 d}) \to \SL(2,\mathbb{Z}_d)$ be the natural
homomorphism defined by
\begin{equation}
 g\colon \begin{pmatrix} \alpha & \beta \\ \gamma & \delta
\end{pmatrix}
\mapsto
\begin{pmatrix} [\alpha]_d & [\beta]_d \\ [\gamma]_d & [\delta]_d
\end{pmatrix}
\end{equation}
where $[x]_d$ denotes the residue class of $x$ \emph{modulo} $d$.   It is easily
seen that $g$ is surjective.  In fact, consider arbitrary 
\begin{equation}
F = \begin{pmatrix} \alpha & \beta \\ \gamma & \delta
\end{pmatrix} \in \SL(2, \mathbb{Z}_d)
\end{equation}
Then $\alpha \delta - \beta \gamma =1 + n d$ for some integer $n$.  If $n$ is
even then $F \in \SL(2, \mathbb{Z}_{\overline{d}})$ and $F=g(F)$.  Suppose, on the
other hand, that $n$ is odd.  Then either $\alpha$ or $\beta$ is odd.  If $\alpha$
is odd $F=g(F')$ where
\begin{equation}
F' = \begin{pmatrix} \alpha & \beta \\ \gamma & \delta + d
\end{pmatrix} \in \SL(2, \mathbb{Z}_{\overline{d}})
\end{equation}
while if $\beta$ is odd $F=g(F'')$ where
\begin{equation}
F'' = \begin{pmatrix} \alpha & \beta \\ \gamma+d & \delta 
\end{pmatrix} \in \SL(2, \mathbb{Z}_{\overline{d}})
\end{equation} 
 Now let
$K_g$ be the kernel of $g$. A matrix $F\in K_g$ if and only if 
\begin{equation}
F=\begin{pmatrix} 1+ r_1 d & s d \\ t d & 1+ r_2 d
\end{pmatrix}
\end{equation}
where $r_1, r_2, s, t=0$ or $1$ and $(1+r_1 d)(1+r_2 d) - s t d^2 = 1\;
(\text{mod}\; 2 d)$.  We have
\begin{equation}
(1+r_1 d)(1+r_2 d) - s t d^2 = 1 + (r_1+ r_2)d \quad (\text{mod} \; 2 d)
\end{equation}
(bearing in mind that $d$ is even, so $d^2=0\; (\text{mod}\; 2d)$).  We
therefore require $r_1=r_2$.  It follows  that
$K_g$ consists of the $8$ matrices of the form 
\begin{equation}
\begin{pmatrix} 1+ r d & s d \\ t d & 1+ r d
\end{pmatrix}
\end{equation}
where $r, s, t= 0$ or $1$. The fact that $g$ is surjective and $|K_g| =8$ implies 
$|\SL(2,\mathbb{Z}_{2 d})|=8|\SL(2,\mathbb{Z}_d)|$.  In view of
Theorem~\ref{thm:CliffordStructure} this means
\begin{equation}
\bigl| \C(d)/\Cc (d)\bigr|=\frac{1}{8}\bigl|\SL(2,
\mathbb{Z}_{2 d})\bigr| \bigl|(\mathbb{Z}_d)^2 \bigr|
=
\bigl|\SL(2,\mathbb{Z}_{ d})\bigr| \bigl|(\mathbb{Z}_d)^2 \bigr|
=
\bigl|\SL(2,
\mathbb{Z}_d)
\ltimes (\mathbb{Z}_d)^2 \bigr|
\end{equation}
as claimed.

We have shown that $\bigl| \C(d)/\Cc (d)\bigr| =\bigl|\SL(2,
\mathbb{Z}_d)
\ltimes (\mathbb{Z}_d)^2 \bigr| = d^2 \bigl| \SL(2,
\mathbb{Z}_d)\bigr| $ for all $d$, odd or even.  It remains to calculate
$\bigl|\SL(2,\mathbb{Z}_d)\bigr|$.  For each $n\in \mathbb{Z}_d$ let $M_n
\subseteq \SL(2,\mathbb{Z}_d)$ be the set of matrices
\begin{equation}
 \begin{pmatrix} \alpha & \beta \\ \gamma & \delta
\end{pmatrix}
\end{equation}
for which $\alpha \delta= n+1 \; (\text{mod}\; d)$ and $\beta \gamma= n \;
(\text{mod}\; d)$.  Clearly  
$
\SL(2,\mathbb{Z}_d)=\bigcup_{n=0}^{d-1}
M_n$
and $|M_n| = \nu(n,d) \nu(n+1,d)$.  It follows that
\begin{equation}
\bigl|\SL(2,\mathbb{Z}_d)\bigr| = \sum_{n=0}^{d-1}\nu(n,d)
\nu(n+1,d)
\end{equation}
Eq.~(\ref{eq:CliffordOrderFormula1}) is now immediate.

If $d$ is a prime number
\begin{equation}
\nu(n,d) =
\begin{cases}
2 d -1 \qquad &\text{if $n=0$ (mod $d$)} \\
d-1 \qquad &\text{otherwise}
\end{cases}
\end{equation}
implying
\begin{equation}
\sum_{n=0}^{d-1}\nu(n,d)
\nu(n+1,d) = d(d^2-1)
\end{equation}
Eq.~(\ref{eq:CliffordOrderFormula2}) is now immediate.
\end{proof}
\section{The Extended Clifford Group}
\label{sec:ExtendedClifford}
It can be seen from Eqs.~(\ref{eq:TDef}--\ref{eq:DOpDefinition})
and~(\ref{eq:GPfiducial}) that, if
$|\psi\rangle = \sum_{r=0}^{d-1} \psi_r |e_r\rangle$ is a GP fiducial vector,
then so is the vector
$|\psi^{*}\rangle = \sum_{r=0}^{d-1} \psi^{*}_r |e_r\rangle$ obtained by complex
conjugation.  So to make the analysis complete we need to consider automorphisms
of $\W(d)$   which are generated by anti-unitary
operators.

An anti-linear operator is a map $\hat{L}\colon \mathbb{C}^d \to \mathbb{C}^d$
with the property
\begin{equation}
 \hat{L} \left(\alpha|\phi\rangle + \beta |\psi\rangle \right)
= \alpha^{*} \hat{L} |\phi \rangle + \beta^{*} \hat{L} |\psi \rangle
\end{equation}
for all $|\phi \rangle, |\psi \rangle \in  \mathbb{C}^d$ and all $\alpha, \beta
\in
\mathbb{C}$.  The adjoint $\hat{L}^{\dagger}$ is defined to be the unique
anti-linear operator  with the property
\begin{equation}
 \langle \phi|\hat{L}^{\dagger} |\psi\rangle = \langle \psi | \hat{L}
|\phi\rangle
\end{equation}
for all $|\phi \rangle, |\psi \rangle \in  \mathbb{C}^d$.  An operator $\hat{U}$
is said to be anti-unitary if it is anti-linear and
$\hat{U}^{\dagger} \hat{U}=1$  (or, equivalently, $\hat{U} \hat{U}^{\dagger} =
1$).

We now define the \emph{extended Clifford Group} to be the group $\EC(d)$
consisting of all  unitary or anti-unitary operators $\hat{U}$ having the
property 
\begin{equation}
 \hat{U} \W(d) \hat{U}^{\dagger} = \W(d)
\end{equation}
 Let us also define
$\ESL(2,\mathbb{Z}_{\overline{d}})$ to be the group consisting of all $2\times 2$
matrices 
\begin{equation}
 \begin{pmatrix}  \alpha & \beta \\ \gamma & \delta
\end{pmatrix}
\end{equation}
such that $\alpha, \beta, \gamma, \delta \in \mathbb{Z}_{\overline{d}}$ and 
$\alpha \delta -\beta \gamma = \pm 1 \;(\text{mod}\; \overline{d})$.  In the last
section we showed that there is a natural  homomorphism
$f\colon \SL(2,\mathbb{Z}_{\overline{d}})\ltimes (\mathbb{Z}_d)^2 \to
\C(d)/\Cc(d)$.  We are going to   show that this extends to a natural 
homomorphism
$f_{\mathrm{E}}\colon\ESL(2,\mathbb{Z}_{\overline{d}})\ltimes (\mathbb{Z}_d)^2 \to
\EC(d)/\Cc(d)$.

Let $\hat{J}$ be the anti-linear operator  which replaces
components in the standard basis with their complex conjugates:
\begin{equation}
\hat{J} \colon \sum_{r=0}^{d-1} \psi_r |e_r\rangle \mapsto
\sum_{r=0}^{d-1} \psi_r^{*} |e_r\rangle
\label{eq:JopDef}
\end{equation}
Clearly $\hat{J}^{\dagger} = \hat{J}$ and $\hat{J}^{\dagger} \hat{J}=\hat{J}^2 =
1$.  So $\hat{J}$ is an anti-unitary operator.  Furthermore, it follows from
Eqs.~(\ref{eq:TDef}--\ref{eq:DOpDefinition}) that 
\begin{equation}
\hat{J}  \hat{D}_{\mathbf{p}} \hat{J}^{\dagger} = \hat{D}_{\tilde{J}\mathbf{p}}
\label{eq:JAction}
\end{equation}
for all $\mathbf{p}$, where 
\begin{equation}
\tilde{J} = \begin{pmatrix} 1 & 0 \\ 0 & -1
\end{pmatrix}
\end{equation}
So $\hat{J}\in \EC(d)$.  Note that $\Det \tilde{J} = -1 \;(\text{mod}\;
\overline{d})$, so $\tilde{J}\in
\ESL(2,\mathbb{Z}_{\overline{d}})$.

Now let $\AC(d)$ be the set of anti-unitary operators $\in \EC(d)$ (so $\EC(d)$
is the disjoint union $\EC(d) =\C(d) \cup \AC(d)$).  The mapping $\hat{U}
\mapsto
\hat{J} \hat{U}$ defines a bijective correspondence between $\AC(d)$ and
$\C(d)$.  We can use this to prove the following extension of
Theorem~\ref{thm:CliffordStructure}:
\begin{theorem}
\label{thm:CliffordStructureE}
There is a unique surjective homomorphism 
\begin{equation}
f_{\mathrm{E}} \colon \ESL (2, \mathbb{Z}_{\overline{d}}) \ltimes (\mathbb{Z}_d)^2
\to
\EC(d)/\Cc(d)
\end{equation}
such  that, for each $(F,\boldsymbol{\chi}) \in
\ESL (2, \mathbb{Z}_{\overline{d}}) \ltimes (\mathbb{Z}_d)^2$
and  $\hat{U} \in
f_{\mathrm{E}} (F,\boldsymbol{\chi})$,
\begin{equation}
\hat{U}
\hat{D}_{\mathbf{p}}
\hat{U}^{\dagger} =\omega^{<\boldsymbol{\chi},F
\mathbf{p}>}\hat{D}_{F\mathbf{p}}
\end{equation}
  for  all
$\mathbf{p}$.  $\hat{U}$ is unitary if 
$\Det F = 1
\;(\text{mod}\;
\overline{d})$ and anti-unitary if  $\Det F = -1 \;(\text{mod}\;
\overline{d})$.

$f_{\mathrm{E}}$ extends the homomorphism $f$ defined in
Theorem~\ref{thm:CliffordStructure}, and has the same kernel.  So
$f_{\mathrm{E}}$ is an isomorphism if $d$ is odd, while if $d$ is even its kernel
is the subgroup $K_f$ defined in Theorem~\ref{thm:CliffordStructure}. 
\end{theorem}

\begin{proof}
Let $\hat{U}$ be an arbitrary anti-unitary operator $\in \AC(d)$.  The fact
that $\hat{J}, \hat{U}$ are both anti-unitary means that $\hat{J}\hat{U}$ is 
unitary.  So $\hat{J}\hat{U}\in \C(d)$.  It then
follows from Theorem~\ref{thm:CliffordStructure} that there exists  $(F',
\boldsymbol{\chi}')\in \SL(2,\mathbb{Z}_{\overline{d}})\ltimes (\mathbb{Z}_d)^2$
such that
\begin{equation}
(\hat{J} \hat{U}) \hat{D}_{\mathbf{p}} (\hat{J}\hat{U})^{\dagger}
=\omega^{<\boldsymbol{\chi}',F' \mathbf{p}>} \hat{D}_{F'\mathbf{p}}
\end{equation}
for all $\mathbf{p}$. Define $F = \tilde{J} F'$ and $\boldsymbol{\chi}=\tilde{J}
\boldsymbol{\chi}'$. In view of Eq.~(\ref{eq:JAction}), and the fact that
$\hat{J}^2=1$, we deduce 
\begin{equation}
\hat{U} \hat{D}_{\mathbf{p}} \hat{U}^{\dagger}
=\hat{J}(\hat{J} \hat{U}) \hat{D}_{\mathbf{p}} (\hat{J}\hat{U})^{\dagger}
\hat{J}^{\dagger}
= \omega^{-<\boldsymbol{\chi}',F' \mathbf{p}>} \hat{D}_{\tilde{J}F'\mathbf{p}}
=\omega^{<\boldsymbol{ \chi},F \mathbf{p}>}
\hat{D}_{F\mathbf{p}}
\end{equation}
for all $\mathbf{p}$ (where we have used the fact that
$<\boldsymbol{\xi},\boldsymbol{\eta}>=-
<\tilde{J}\boldsymbol{\xi},\tilde{J}\boldsymbol{\eta}>$ for all
$\boldsymbol{\xi}, \boldsymbol{\eta}$).  We have
$\Det (F)=(\Det
\bar{J})(\Det F')=-1$, so $(F,
\boldsymbol{\chi})\in\ESL(2,\mathbb{Z}_{\overline{d}})\ltimes
(\mathbb{Z}_d)^2$. 

Reversing the argument we deduce the converse proposition:  for each $(F,
\boldsymbol{\chi})\in
\ESL(2,\mathbb{Z}_{\overline{d}})\ltimes (\mathbb{Z}_d)^2$, there exists $\hat{U}
\in \EC(d)$ such that  $\hat{U} \hat{D}_{\mathbf{p}}
\hat{U}^{\dagger}=\omega^{<\boldsymbol{ \chi},F
\mathbf{p}>}\hat{D}_{F\mathbf{p}}$ for all
$\mathbf{p}$.  The fact that an operator commutes with
$\hat{D}_{\mathbf{p}}$ for all $\mathbf{p}$ if and only if it is a multiple of
the identity means that $\hat{U}$ is unique up to a phase. 

This establishes the existence and uniqueness of the homomorphism
$f_{\mathrm{E}}$.  The proof of the remaining statements is straightforward, and
is left to the reader.
\end{proof}
Finally, we have the following result which, together with
Lemma~\ref{lem:CliffordStructure5}, enables us to calculate the order of 
$\EC(d)/\Cc(d)$:
\begin{lemma}
\label{lem:OrderECd}
\begin{equation}
 \bigl| \EC(d)/\Cc(d)  \bigr|=2\bigl| \C(d)/\Cc(d)  \bigr|
\end{equation}
for all $d$.
\end{lemma}
\begin{proof}
The map 
\begin{equation}
\hat{U} \Cc(d)\mapsto \hat{J} \hat{U} \Cc(d)
\end{equation}
defines a bijective correspondence between
$\AC(d) /\Cc(d)$  and $\C(d)/\Cc(d)$.  So the set  $\AC(d) /\Cc(d)$ contains the
same number of elements as $\C(d) /\Cc(d)$.  The statement is now immediate.
\end{proof}

\section{The Clifford Trace}
\label{sec:CliffordTrace}
We now define the Clifford trace.  The significance of this function for us is
that every GP fiducial vector which has been constructed to
date
 is an eigenvector of a Clifford unitary having Clifford trace
$=-1$.

 Let $[F,
\boldsymbol{\chi}]\in
\EC(d)/
\Cc(d)$ be the image of
$(F,\boldsymbol{\chi})$ under the homomorphism $f_{\mathrm{E}}$ defined in 
Theorem~\ref{thm:CliffordStructureE}.    We refer to $[F,
\boldsymbol{\chi}]$ as an extended Clifford operation (or Clifford
operation if it $\in \C(d)/\Cc(d)$).  The operators $\in [F,
\boldsymbol{\chi}]$ only differ by a phase.   It is therefore convenient to adopt
a terminology which blurs the distinction between the operation $[F,
\boldsymbol{\chi}]$ and the operators $\hat{U}\in [F,
\boldsymbol{\chi}]$.  In particular, we will adopt the 
convention that properties which hold for each
$\hat{U}\in [F,\boldsymbol{\chi}]$ may also be attributed to $[F,
\boldsymbol{\chi}]$.  Thus, we
will say that
$[F,
\boldsymbol{\chi}]$ is unitary (respectively anti-unitary) if the operators
$\hat{U}\in [F,
\boldsymbol{\chi}]$ are unitary (respectively  anti-unitary).  Similarly, we will
say that $|\psi\rangle \in \mathbb{C}^d$ is an eigenvector of $[F,
\boldsymbol{\chi}]$ if it is an eigenvector of the operators
$\hat{U}\in [F,
\boldsymbol{\chi}]$.

It is easily verified that  $\Tr (F_1) =
\Tr(F_2)\;(\text{mod}\;d)$ whenever $[F_1,
\boldsymbol{\chi}_1]=[F_2,\boldsymbol{\chi}_2]$  (note that it is not necessarily
true that $\Tr (F_1) =
\Tr(F_2)\;(\text{mod}\;\overline{d})$ if $d$ is even).  We  therefore obtain a
well-defined function $\EC(d)/\Cc(d) \to \mathbb{Z}_d$ if we assign to each
operation $[F,\boldsymbol{\chi}]$ the value $\Tr(F)\;(\text{mod}\;d)$.  We obtain
a function $\EC(d) \to \mathbb{Z}_d$ by assigning to each $\hat{U} \in
[F,\boldsymbol{\chi}]$ the value $\Tr(F)\;(\text{mod}\;d)$.  We  use the term
``Clifford trace'' to refer to either of these functions.

We now prove the main result of this section, which states that there is a
connection between the order of a Clifford operation and its Clifford trace.
\begin{lemma}
\label{lem:CliffordTrace}
Let $[F,\boldsymbol{\chi}] \in
\C(d)/\Cc(d)$, where $d$ is any dimension $\neq 3$. Then
$[F,\boldsymbol{\chi}]$ is of order
$3$ if $\Tr (F)=-1\;(\text{mod}\;d)$.

Let $[F,\boldsymbol{\chi}] \in
\C(d)/\Cc(d)$, where $d$ is any prime dimension $\neq 3$. Then the stronger
statement is true:
$[F,\boldsymbol{\chi}]$ is of order
$3$ if and only if $\Tr (F)=-1\;(\text{mod}\;d)$.
\end{lemma}
\begin{remark}
The restriction to  operations $\in \C(d)/\Cc(d)$ is essential (because if
$[F,\boldsymbol{\chi}]$ is anti-unitary its order must be even).
\end{remark}

\begin{proof}
Let  $[F,
\boldsymbol{\chi}]\in \C(d)/\Cc(d)$, and let $\kappa = \Tr (F)$.  Then, taking
into account the fact that $\Det (F)=1 \; (\text{mod}\; \overline{d})$, it is
straightforward to show
\begin{align}
F^2 & = \kappa F -1 & & (\text{mod}\; \overline{d}) 
\label{eq:F2TermsTraceF}
\\
\intertext{implying}
 F^3  & = (\kappa^2 -1) F - \kappa &  &(\text{mod}\; \overline{d}) 
\label{eq:F3TermsTraceF}
\\
1+F + F^2 & = (\kappa +1) F &  &(\text{mod}\; \overline{d}) 
\end{align}
Now suppose that $\kappa = -1 \;(\text{mod}\; d)$.  Then there are three
possibilities:  (a) $d$ is odd; (b) $d$ is even and $\kappa = -1 \;(\text{mod}\;
\overline{d})$; (c) $d$ is even and $\kappa = -1+d \;(\text{mod}\;
\overline{d})$.  In case (a) or (b) we have
\begin{align}
 F^3  & = 1 &  &(\text{mod}\; \overline{d}) 
\\
1+F + F^2 & = 0 &  &(\text{mod}\; \overline{d}) 
\end{align}
while in case (c) we have $\kappa^2-1 = d^2 - 2 d = 0\;(\text{mod}\;
\overline{d})$, and consequently
\begin{align}
F^3 & = \begin{pmatrix} 1+ d & 0 \\ 0 & 1+ d \end{pmatrix} 
&  &(\text{mod}\; \overline{d}) 
\\
1+F + F^2 & = 0 &  &(\text{mod}\; d) 
\end{align}
Referring to the definition of $K_f$ (see Theorem~\ref{thm:CliffordStructure}) we
deduce that, in every case,
\begin{equation}
(F,\boldsymbol{\chi})^3=(F^3,(1+F+F^2)\boldsymbol{\chi}) \in K_f
\end{equation}
implying that $[F,
\boldsymbol{\chi}]^3=[1,\boldsymbol{0}]$.
It remains to show that neither $[F,
\boldsymbol{\chi}]$ nor  $[F,
\boldsymbol{\chi}]^2=[1,\boldsymbol{0}]$.  To see that
$[F,
\boldsymbol{\chi}] \neq [1,\boldsymbol{0}]$ observe that the contrary would imply 
$-1 = \kappa = \Tr({1}) = 2 \;(\text{mod}\; d)$, which is not possible
given that $d \neq 3$.  Similarly,
if $[F,
\boldsymbol{\chi}]^2=[1,\boldsymbol{0}]$ it would follow (taking the trace on
both sides of Eq.~(\ref{eq:F2TermsTraceF})) that $2 = \kappa^2-2=-1
\;(\text{mod}\; d)$, contrary to the assumption that $d\neq 3$.  We conclude that 
$[F,
\boldsymbol{\chi}]$ is of order $3$, as claimed.

To prove the second part of the lemma suppose that $d$ is a prime number $\neq 3$
and $[F,
\boldsymbol{\chi}]$ is of order $3$.  Then $(F^3,(1+F+F^2)\boldsymbol{\chi}) \in
K_f$, implying
$F^3 = 1 \; (\text{mod} \; d)$.  In view of Eq.~(\ref{eq:F3TermsTraceF}) this
means
\begin{equation}
(\kappa+1)\left( (\kappa-1) F -1\right) = 0 \qquad (\text{mod}\; d)
\label{eq:F3Equal1Consequence}
\end{equation}
We now proceed by \emph{reductio ad absurdum}.  Suppose that $\kappa \neq
-1\;(\text{mod}\; d)$. Then Eq.~(\ref{eq:F3Equal1Consequence}) and the fact that
$d$ is prime  implies
\begin{equation}
(\kappa-1) F =1\qquad (\text{mod}\; d)
\label{eq:F3Equal1ConsequenceB}
\end{equation}
Taking the trace on both sides gives $(\kappa+1)(\kappa -2) =
0\;(\text{mod}\; d) $ implying $\kappa = 2\;(\text{mod}\; d) $.  Substituting
this value
 into Eq.~(\ref{eq:F3Equal1ConsequenceB}) we deduce $F = 1\;(\text{mod}\; d)$,
implying $F^2 = 1\;(\text{mod}\; \overline{d})$ and $F^3 = F\;(\text{mod}\;
\overline{d})$.  So
\begin{equation}
(F, 3\boldsymbol{\chi})
=(F,\boldsymbol{\chi})^3 \in  K_f
\end{equation}
implying $(F, \boldsymbol{\chi})\in  K_f$.  But that would mean
$[F,
\boldsymbol{\chi}]$ is of order $1$, contrary to assumption.
  We conclude that $\kappa =-1
\;(\text{mod}\; d)$, as claimed.
\end{proof}
The result does not hold when $d=3$ because then the identity has Clifford trace
$=-1$.  It is, however, easily verified that in dimension $3$ (as in every other
prime dimension) every order $3$ Clifford operation has Clifford trace $=-1$.

If $d$ is not a prime number there may exist order $3$ Clifford operations for
which the Clifford trace $\neq-1$.  Consider, for example,
\begin{equation}
[F,\boldsymbol{\chi}]=\left[\begin{pmatrix} 5 & 4 \\ 2 & -3
\end{pmatrix},
\begin{pmatrix} - 4 \\ 5
\end{pmatrix}
\right] \in \C(6)/\Cc(6)
\end{equation}
Then $[F,\boldsymbol{\chi}]$ is of order $3$ yet $\Tr(F) =2 \;
(\text{mod}\; 6)$.  

Because these results will play an important role in the following  it is
convenient to introduce some terminology.  We will say that an operation
$[F,\boldsymbol{\chi}]\in \C(d)/\Cc(d)$ is a \emph{canonical}  order $3$ unitary
if
\begin{enumerate}
\item[(a)] $\Tr (F)=-1\;(\text{mod} \; d)$.
\item[(b)] $F$ is not the identity matrix.
\end{enumerate}
Note that the second stipulation is only needed because of the possibility
that $d=3$.  If $d\neq 3$ an operation $[F,\boldsymbol{\chi}]\in
\C(d)/\Cc(d)$ is a  canonical order $3$ unitary if and only if $\Tr
(F)=-1\;(\text{mod}
\; d)$.  
\section{The RBSC Vectors}
\label{sec:RBSCVectors}
For  $5\le d \le 45$ 
RBSC~\cite{Renes,RenesVectors} have constructed GP fiducial vectors numerically. 
In this section we examine the behaviour of these vectors under the action of the
extended Clifford group.  In particular we show that each of them is an
eigenvector of a canonical order $3$ Clifford unitary.  This suggests 
\begin{quote}
\textbf{Conjecture A:}  GP fiducial vectors exist in every finite dimension.  
Furthermore, every such vector  is an eigenvector of a
canonical order
$3$ unitary.
\end{quote}

Conjecture A is related to a conjecture of Zauner's.  Let
\begin{equation}
[ Z,\boldsymbol{0}]=\left[
\begin{pmatrix} 0 & -1 \\ 1 & -1
\end{pmatrix},
\begin{pmatrix} 0 \\ 0
\end{pmatrix}
\right]
\end{equation}
It will be observed that $[ Z,\boldsymbol{0}]$ is defined, and $\in
\C(d)/\Cc(d)$, for every dimension
$d$, and that it is canonical order
$3$.  Zauner~\cite{Zauner} has conjectured
\begin{quote}
\textbf{Conjecture B:}  In each dimension $d$ there exists a GP fiducial vector
which is an eigenvector of $[ Z,\boldsymbol{0}]$.
\end{quote}
In Section~\ref{sec:Zauner} we will see that RBSC's numerical data also provides
further support for Conjecture B.

Let $|\psi_d\rangle$ be the RBSC vector in dimension $d$.  In
Table~\ref{tbl:RBSC} we list, for each value of $d$, a unitary Clifford operation
$[F_d,
\boldsymbol{\chi}_d]$ having $|\psi_d\rangle$ as one of its eigenvectors. It will
be seen that, in every case, $\Tr(F_d) = -1 \;(\text{mod}\;d)$,
implying that
$[F_d,
\boldsymbol{\chi}_d]$ is canonical order $3$.  Clearly, 
$|\psi_d\rangle$ is also an eigenvector of $[F_d,
\boldsymbol{\chi}_d]^2$.  Moreover, $[F_d,
\boldsymbol{\chi}_d]^2$ also has Clifford trace $=-1$.  There are, however, no
other Clifford operations with these properties.

 In Table~\ref{tbl:RBSC} we also list  $(n_{d1},n_{d2},n_{d3})$,
the dimensions of the three eigenspaces of $[F_d,
\boldsymbol{\chi}_d]$, and $n_d$, the dimension of the
particular eigenspace to which $|\psi_d\rangle$ belongs.  It will be seen that,
with one exception, $|\psi_d\rangle$ always belongs to an eigenspace of highest
dimension (the exception being $d=17$, where $|\psi_d\rangle$ belongs to the
eigenspace of
\emph{lowest} dimension).
\begin{table}
\SMALL
\begin{center}
\begin{tabular}{|c c c c c||c c c c c|}
\hline
\parbox{1 pt}{\rule{0 ex}{5 ex}}
$d$ & $F_d$ & $\boldsymbol{\chi}_d$ & $(n_{d1},n_{d2},n_{d3})$ & $n_d$ \hspace{1
ex} &
\hspace{1 ex}
$d$ &
$F_d$ &
$\boldsymbol{\chi}_d$ & $(n_{d1},n_{d2},n_{d3})$ & $n_d$
\\
\hline 
    \myStrut $5$  &  
        $\begin{pmatrix}  -1  &  -1  \\  1  &  0  \end{pmatrix}$  &  
        $\begin{pmatrix}  2  \\  2  \end{pmatrix}$  &  
        $(1, 2, 2)$  &  $2$  
        \hspace{1 ex}  &  \hspace{1 ex}    
    \myStrut $26$  &  
        $\begin{pmatrix}  -7  &  -9  \\  -1  &  6  \end{pmatrix}$  &  
        $\begin{pmatrix}  -11  \\  11  \end{pmatrix}$  &  
        $(8, 9, 9)$  &  $9$  
\\  \myStrut $6$  &  
        $\begin{pmatrix}  -2  &  3  \\  -1  &  1  \end{pmatrix}$  &  
        $\begin{pmatrix}  3  \\  0  \end{pmatrix}$  &  
        $(1, 2, 3)$  &  $3$  
        \hspace{1 ex}  &   \hspace{1 ex}    
    \myStrut $27$  &  
        $\begin{pmatrix}  -10  &  1  \\  -10  &  9  \end{pmatrix}$  &  
        $\begin{pmatrix}  -3  \\  -12  \end{pmatrix}$  &  
        $(8, 9, 10)$  &  $10$  
\\  \myStrut $7$  &  
        $\begin{pmatrix}  -2  &  -2  \\  -2  &  1  \end{pmatrix}$  &  
        $\begin{pmatrix}  2  \\  0  \end{pmatrix}$  &  
        $(2, 2, 3)$  &  $3$  
        \hspace{1 ex}  &  \hspace{1 ex}    
    \myStrut $28$  &  
        $\begin{pmatrix}  -3  &  21  \\  5  &  2  \end{pmatrix}$  &  
        $\begin{pmatrix}  -10  \\  -6  \end{pmatrix}$  &  
        $(9, 9, 10)$  &  $10$  
\\  \myStrut $8$  &  
        $\begin{pmatrix}  -4  &  3  \\  1  &  3  \end{pmatrix}$  &  
        $\begin{pmatrix}  3  \\  -1  \end{pmatrix}$  &  
        $(2, 3, 3)$  &  $3$  
        \hspace{1 ex}  &   \hspace{1 ex}    
    \myStrut $29$  &  
        $\begin{pmatrix}  -13  &  -6  \\  2  &  12  \end{pmatrix}$  &  
        $\begin{pmatrix}  -10  \\  12  \end{pmatrix}$  &  
        $(9, 10, 10)$  &  $10$  
\\  \myStrut $9$  &  
        $\begin{pmatrix}  -3  &  2  \\  1  &  2  \end{pmatrix}$  &  
        $\begin{pmatrix}  2  \\  1  \end{pmatrix}$  &  
        $(2, 3, 4)$  &  $4$  
        \hspace{1 ex}  &   \hspace{1 ex}    
    \myStrut $30$  &  
        $\begin{pmatrix}  -8  &  -7  \\  -9  &  7  \end{pmatrix}$  &  
        $\begin{pmatrix}  11  \\  -3  \end{pmatrix}$  &  
        $(9, 10, 11)$  &  $11$  
\\  \myStrut $10$  &  
        $\begin{pmatrix}  -4  &  -7  \\  -1  &  3  \end{pmatrix}$  &  
        $\begin{pmatrix}  -2  \\  0  \end{pmatrix}$  &  
        $(3, 3, 4)$  &  $4$  
        \hspace{1 ex}  &  \hspace{1 ex}    
    \myStrut $31$  &  
        $\begin{pmatrix}  -9  &  -10  \\  -2  &  8  \end{pmatrix}$  &  
        $\begin{pmatrix}  -14  \\  6  \end{pmatrix}$  &  
        $(10, 10, 11)$  &  $11$  
\\  \myStrut $11$  &  
        $\begin{pmatrix}  -5  &  4  \\  3  &  4  \end{pmatrix}$  &  
        $\begin{pmatrix}  -5  \\  0  \end{pmatrix}$  &  
        $(3, 4, 4)$  &  $4$  
        \hspace{1 ex}  &    \hspace{1 ex}    
    \myStrut $32$  &  
        $\begin{pmatrix}  -11  &  -31  \\  -15  &  10  \end{pmatrix}$  &  
        $\begin{pmatrix}  11  \\  -7  \end{pmatrix}$  &  
        $(10, 11, 11)$  &  $11$  
\\  \myStrut $12$  &  
        $\begin{pmatrix}  -4  &  11  \\  1  &  3  \end{pmatrix}$  &  
        $\begin{pmatrix}  4  \\  -5  \end{pmatrix}$  &  
        $(3, 4, 5)$  &  $5$  
        \hspace{1 ex}  &   \hspace{1 ex}    
    \myStrut $33$  &  
        $\begin{pmatrix}  -7  &  -5  \\  2  &  6  \end{pmatrix}$  &  
        $\begin{pmatrix}  8  \\  -5  \end{pmatrix}$  &  
        $(10, 11, 12)$  &  $12$  
\\ \myStrut $13$  &  
        $\begin{pmatrix}  -2  &  -2  \\  -5  &  1  \end{pmatrix}$  &  
        $\begin{pmatrix}  6  \\  0  \end{pmatrix}$  &  
        $(4, 4, 5)$  &  $5$  
        \hspace{1 ex}  &  \hspace{1 ex}    
   \myStrut $34$  &  
        $\begin{pmatrix}  -12  &  3  \\  1  &  11  \end{pmatrix}$  &  
        $\begin{pmatrix}  -1  \\  -16  \end{pmatrix}$  &  
        $(11, 11, 12)$  &  $12$  
\\ \myStrut $14$  &  
        $\begin{pmatrix}  -2  &  -3  \\  1  &  1  \end{pmatrix}$  &  
        $\begin{pmatrix}  -5  \\  1  \end{pmatrix}$  &  
        $(4, 5, 5)$  &  $5$  
        \hspace{1 ex}  &   \hspace{1 ex}    
   \myStrut $35$  &  $\begin{pmatrix}  -13  &  -12  \\  16  &  12  \end{pmatrix}$  &  
        $\begin{pmatrix}  11  \\  -12  \end{pmatrix}$  &  
        $(11, 12, 12)$  &  $12$  
\\ \myStrut $15$  &  
        $\begin{pmatrix}  -5  &  1  \\  -6  &  4  \end{pmatrix}$  &  
        $\begin{pmatrix}  -7  \\  -6  \end{pmatrix}$  &  
        $(4, 5, 6)$  &  $6$  
        \hspace{1 ex}  &   \hspace{1 ex}    
   \myStrut $36$  &  
        $\begin{pmatrix}  -8  &  21  \\  -13  &  7  \end{pmatrix}$  &  
        $\begin{pmatrix}  0  \\  7  \end{pmatrix}$  &  
        $(11, 12, 13)$  &  $13$  
\\ \myStrut $16$  &  
        $\begin{pmatrix}  -8  &  13  \\  3  &  7  \end{pmatrix}$  &  
        $\begin{pmatrix}  1  \\  0  \end{pmatrix}$  &  
        $(5, 5, 6)$  &  $6$  
        \hspace{1 ex}  &  \hspace{1 ex}    
   \myStrut $37$  &  
        $\begin{pmatrix}  -16  &  1  \\  18  &  15  \end{pmatrix}$  &  
        $\begin{pmatrix}  -4  \\  3  \end{pmatrix}$  &  
        $(12, 12, 13)$  &  $13$  
\\ \myStrut $17$  &  
        $\begin{pmatrix}  -5  &  -7  \\  3  &  4  \end{pmatrix}$  &  
        $\begin{pmatrix}  6  \\  7  \end{pmatrix}$  &  
        $(5, 6, 6)$  &  $5$  
        \hspace{1 ex}  &  \hspace{1 ex}    
   \myStrut $38$  &  
        $\begin{pmatrix}  -6  &  -31  \\  1  &  5  \end{pmatrix}$  &  
        $\begin{pmatrix}  12  \\  -10  \end{pmatrix}$  &  
        $(12, 13, 13)$  &  $13$  
\\ \myStrut $18$  &  
        $\begin{pmatrix}  -5  &  5  \\  3  &  4  \end{pmatrix}$  &  
        $\begin{pmatrix}  9  \\  0  \end{pmatrix}$  &  
        $(5, 6, 7)$  &  $7$  
        \hspace{1 ex}  &  \hspace{1 ex}    
   \myStrut $39$  &  
        $\begin{pmatrix}  -17  &  -11  \\  0  &  16  \end{pmatrix}$  &  
        $\begin{pmatrix}  8  \\  15  \end{pmatrix}$  &  
        $(12, 13, 14)$  &  $14$  
\\ \myStrut $19$  &  
        $\begin{pmatrix}  -2  &  4  \\  4  &  1  \end{pmatrix}$  &  
        $\begin{pmatrix}  -7  \\  -4  \end{pmatrix}$  &  
        $(6, 6, 7)$  &  $7$  
        \hspace{1 ex}  &  \hspace{1 ex}    
   \myStrut $40$  &  
        $\begin{pmatrix}  -3  &  19  \\  -13  &  2  \end{pmatrix}$  &  
        $\begin{pmatrix}  -12  \\  -19  \end{pmatrix}$  &  
        $(13, 13, 14)$  &  $14$  
\\ \myStrut $20$  &  
        $\begin{pmatrix}  -2  &  -3  \\  1  &  1  \end{pmatrix}$  &  
        $\begin{pmatrix}  -9  \\  -6  \end{pmatrix}$  &  
        $(6, 7, 7)$  &  $7$  \hspace{1 ex}  &  \hspace{1 ex}    
   \myStrut $41$  &  
        $\begin{pmatrix}  -2  &  -10  \\  -12  &  1  \end{pmatrix}$  &  
        $\begin{pmatrix}  19  \\  13  \end{pmatrix}$  &  
        $(13, 14, 14)$  &  $14$  
\\ \myStrut $21$  &  
        $\begin{pmatrix}  -5  &  -6  \\  -7  &  4  \end{pmatrix}$  &  
        $\begin{pmatrix}  -6  \\  1  \end{pmatrix}$  &  
        $(6, 7, 8)$  &  $8$  
        \hspace{1 ex}  &  \hspace{1 ex}    
   \myStrut $42$  &  
        $\begin{pmatrix}  -15  &  11  \\  19  &  14  \end{pmatrix}$  &  
        $\begin{pmatrix}  0  \\  -15  \end{pmatrix}$  &  
        $(13, 14, 15)$  &  $15$  
\\ \myStrut $22$  &  
        $\begin{pmatrix}  -2  &  -1  \\  3  &  1  \end{pmatrix}$  &  
        $\begin{pmatrix}  8  \\  2  \end{pmatrix}$  &  
        $(7, 7, 8)$  &  $8$  \hspace{1 ex}  &  \hspace{1 ex}    
   \myStrut $43$  &  
        $\begin{pmatrix}  -11  &  1  \\  18  &  10  \end{pmatrix}$  &  
        $\begin{pmatrix}  -1  \\  21  \end{pmatrix}$  &  
        $(14, 14, 15)$  &  $15$  
\\ \myStrut $23$  &  
        $\begin{pmatrix}  -11  &  -10  \\  -5  &  10  \end{pmatrix}$  &  
        $\begin{pmatrix}  0  \\  -3  \end{pmatrix}$  &  
        $(7, 8, 8)$  &  $8$  
        \hspace{1 ex}  &  \hspace{1 ex}    
   \myStrut $44$  &  
        $\begin{pmatrix}  -8  &  -29  \\  5  &  7  \end{pmatrix}$  &  
        $\begin{pmatrix}  16  \\  -5  \end{pmatrix}$  &  
        $(14, 15, 15)$  &  $15$  
\\ \myStrut $24$  &  
        $\begin{pmatrix}  -2  &  -3  \\  1  &  1  \end{pmatrix}$  &  
        $\begin{pmatrix}  0  \\  -3  \end{pmatrix}$  &  
        $(7, 8, 9)$  &  $9$  
        \hspace{1 ex}  &  \hspace{1 ex}    
   \myStrut $45$  &  
        $\begin{pmatrix}  -20  &  -1  \\  21  &  19  \end{pmatrix}$  &  
        $\begin{pmatrix}  -8  \\  6  \end{pmatrix}$  &  
        $(14, 15, 16)$  &  $16$  
\\ \myStrut $25$  &  
        $\begin{pmatrix}  -6  &  -1  \\  6  &  5  \end{pmatrix}$  &  
        $\begin{pmatrix}  -7  \\  12  \end{pmatrix}$  &  
        $(8, 8, 9)$  &  $9$  
        \hspace{1 ex}  &  \hspace{1 ex}     
   \myStrut  &  &  &  &  \\ 
\hline
\end{tabular}
\normalsize
\vspace{2 ex}
\caption{For each $d$ the RBSC vector $|\psi_d\rangle$ is an eigenvector of the
unitary operation $[F_d, \boldsymbol{\chi}_d]$.  Note that in every case $\Tr F_d
=-1$, implying that $[F_d, \boldsymbol{\chi}_d]$ is canonical order $3$.
$(n_{d1},n_{d2},n_{d3})$ are the dimensions of the three eigenspaces of
$[F_d, \boldsymbol{\chi}_d]$, and $n_d$ is the dimension of the eigenspace to
which $|\psi_d\rangle$ belongs.  Note that $n_d = \max (n_{d1},n_{d2},n_{d3})$,
with the single exception of
$d=17$.}
\label{tbl:RBSC}
\end{center}
\end{table}

We used a computer algebra package (\emph{Mathematica}) to construct the table.
To illustrate the method employed  we give a detailed
description for the case $d=5$.   We begin with the observation that, if
$|\psi_5\rangle$ is an eigenvector of $[F,\boldsymbol{\chi}]$, then
\begin{equation}
\langle \psi_5 | \hat{D}_{\mathbf{p}} |\psi_5\rangle
=e^{\frac{2 \pi i}{5}\langle \boldsymbol{\chi},F \mathbf{p}\rangle}
\langle \psi_5 | \hat{D}_{F\mathbf{p}} |\psi_5\rangle
\end{equation} 
for all $\mathbf{p}$. So, using the value of $|\psi_5\rangle$ which is available
on RBSC's website~\cite{RenesVectors}, we look for values of
$\mathbf{p}$,
$\mathbf{q}$ such that
\begin{equation}
\frac{5}{2 \pi}\left( \arg\left(\langle \psi_5 | \hat{D}_{\mathbf{p}}
|\psi_5\rangle  \right)-\arg\left(\langle \psi_5 | \hat{D}_{\mathbf{q}}
|\psi_5\rangle  \right)
\right)
\end{equation}
is an (approximate) integer.  We find that if $\mathbf{p} =(1,0) $ this is only
true when $\mathbf{q} = (1,0), (-1,1)$ or $(0,-1)\;(\text{mod}\;5)$, and that if 
$\mathbf{p} =(0,1) $ it is only true when 
$\mathbf{q} = (0,1), (-1,0)$ or $(1,-1)\;(\text{mod}\;5)$.  Taking  account of
the requirement $\Det(F)=1 \;(\text{mod}\;5)$ we deduce that the only candidates
are (apart from the identity)
\begin{equation}
[F_5,\boldsymbol{\chi}_5]=
\left[\begin{pmatrix} -1 & -1 \\ 1 & 0
\end{pmatrix},
\begin{pmatrix} 2 \\ 2
\end{pmatrix}\right]
\end{equation}
and its square, $[F_5,\boldsymbol{\chi}_5]^2$.  To check that $|\psi_5\rangle$
actually is an eigenvector of $[F_5,\boldsymbol{\chi}_5]$ we observe that $F_5$ is
a prime  matrix.  So in view of Lemma~\ref{lem:CliffordStructure4} we have the
following explicit formula for the
$\hat{U}\in[F_5,\boldsymbol{\chi}_5]$:
\begin{equation}
\hat{U} =\frac{1}{\sqrt{5}} e^{i \theta} \hat{D}_{(2,2)}
\left(\sum_{r,s=0}^{4} e^{-\frac{4 \pi i}{5}s(s+2 r)} |e_r\rangle \langle e_s |
\right)
\end{equation}
 $e^{i \theta}$ being an arbitrary phase. Suppose we choose $\theta=\frac{7 \pi
}{15}$.  Then we find $\hat{U}^3=1$ and 
\begin{equation}
\bigl\| (\hat{U} -1)|\psi_{5} \rangle \bigr\|^{2}= 0
\end{equation}
to machine precision.  This confirms that $|\psi_5\rangle$ is indeed an
eigenvector of 
$[F_5,\boldsymbol{\chi}_5]$.  To calculate the dimensions of the eigenspaces
define, for $r = 0, \pm 1$ (and with the same choice of $\theta$),
\begin{equation}
 \hat{P}_r = \frac{1}{3} \left(1+ e^{-\frac{2 r \pi i}{3}} \hat{U} + 
e^{\frac{2 r \pi i}{3}} \hat{U}^2
\right)
\end{equation}
Then $\hat{P}_r$ projects onto the eigenspace of $\hat{U}$ with eigenvalue 
$e^{\frac{2 r \pi i}{3}}$.  We find
\begin{equation}
\Tr(\hat{P_r}) =
\begin{cases} 1 \qquad & r=1  \\
2 \qquad & r = -1 \;\text{or}\; 0
\end{cases}
\end{equation}
implying that the dimensions of the eigenspaces are $1, 2, 2$, and that
$|\psi_{5}\rangle$ is in one of the eigenspaces with dimension $2$.

In dimensions $6$ to $45$ the calculation goes through in essentially the same
way.  The calculation is, however, slightly more complicated when $d$ is
even, due to the fact that we must then require $\Det F_d=1\;(\text{mod}\; 2 d)$. 
Note, also, that when $d=6, 21, 24, 28$ or $36$ the matrix $F_d$ is non-prime, so
we have to use the decomposition of Lemma~\ref{lem:CliffordStructure3}.

This method also enables us to establish the full stability group of
$|\psi_d\rangle$: 
\emph{i.e.}~the set of all operations (unitary or anti-unitary) $\in
\EC(d)/\Cc(d)$ of which $|\psi_d\rangle$ is an eigenvector.  It turns out that,
with one exception, the stability group is the order $3$ cyclic subgroup
generated by $[F_d,\boldsymbol{\chi}_{d}]$.  The
exception is dimension~$7$, where the stability group is the order~$6$ cyclic
subgroup generated by the anti-unitary operation
\begin{equation}
\left[ A_{\mathstrut 7}, \boldsymbol{\xi}_7
\right]
=
\left[ \begin{pmatrix} 2 & -1 \\ -1 & 0
\end{pmatrix}, 
\begin{pmatrix} 1\\ 1
\end{pmatrix}
\right]
\end{equation}
Note that $[ A^{\mathstrut}_{ 7}, \boldsymbol{\xi}_7 
]^2=[ F^{\mathstrut}_{ 7}, \boldsymbol{\chi}_{7}]$.

\section{Zauner's Conjecture}
\label{sec:Zauner} 
In the last section we saw that RBSC's numerical results
support Conjecture A.  Their results also support Conjecture B: 
\emph{i.e.}\ Zauner's conjecture, that in each dimension $d$ there exists a
GP fiducial vector which is an eigenvector of $[Z,\boldsymbol{0}]$.

In fact, for each $5\le d \le 45$ let $[L_d,\boldsymbol{\eta}_d]$ be the
operation specified in Table~\ref{tbl:Conjugacy}.
\begin{table}
\SMALL
\begin{center}
\begin{tabular}{|c c c ||c c c || c c c|}
\hline
\parbox{1 pt}{\rule{0 ex}{5 ex}}
$d$ & $L_d$ & $\boldsymbol{\eta}_d$  \hspace{1
ex} &
\hspace{1 ex}
$d$ &
$L_d$ &
$\boldsymbol{\eta}_d$ \hspace{1 ex} &
\hspace{1 ex} 
$d$ &
$L_d$ &
$\boldsymbol{\eta}_d$ \\
\hline
 \myStrut
$5$   & $\begin{pmatrix}  1   &   0   \\  1    &   1    \end{pmatrix}$  
      & $\begin{pmatrix}  0   \\  -2                    \end{pmatrix}$ &
$19$  & $\begin{pmatrix}  2   &   1   \\  0    &   -9   \end{pmatrix}$  
      & $\begin{pmatrix}  5   \\  -6                    \end{pmatrix}$  &
$33$  & $\begin{pmatrix}  6   &   2   \\  5    &   -15  \end{pmatrix}$ 
      & $\begin{pmatrix}  13  \\  15                    \end{pmatrix}$  \\
 \myStrut
$6$   & $\begin{pmatrix}  0   &   1   \\  1    &   -1   \end{pmatrix}$  
      & $\begin{pmatrix}  1   \\  -1                    \end{pmatrix}$  &
$20$  & $\begin{pmatrix}  0   &   1   \\ -1    &   -1   \end{pmatrix}$  
      & $\begin{pmatrix}  9   \\  -3                    \end{pmatrix}$  &
$34$  & $\begin{pmatrix}  0   &   1   \\ -1    &   -11  \end{pmatrix}$  
      & $\begin{pmatrix}  13  \\  3                     \end{pmatrix}$  \\
 \myStrut
$7$   & $\begin{pmatrix}  2   &   0   \\  -3   &   -3   \end{pmatrix}$  
      & $\begin{pmatrix}  0   \\  3                     \end{pmatrix}$  &
$21$  & $\begin{pmatrix}  2   &   1   \\  -4   &   8    \end{pmatrix}$  
      & $\begin{pmatrix}  -3  \\  -7                    \end{pmatrix}$  &
$35$  & $\begin{pmatrix}  14  &   2   \\  10   &   4    \end{pmatrix}$  
      & $\begin{pmatrix}  4   \\  6                     \end{pmatrix}$  \\
 \myStrut
$8$   & $\begin{pmatrix}  0   &   1   \\   -1  &   -3   \end{pmatrix}$  
      & $\begin{pmatrix}  -2  \\  3                     \end{pmatrix}$  &
$22$  & $\begin{pmatrix}  1   &   0   \\   2   &   1    \end{pmatrix}$  
      & $\begin{pmatrix}  8   \\  6                     \end{pmatrix}$  &
$36$  & $\begin{pmatrix}  17  &   1   \\   5   &   -4   \end{pmatrix}$  
      & $\begin{pmatrix}  -2  \\  -5                    \end{pmatrix}$ \\
 \myStrut
$9$   & $\begin{pmatrix}  2   &   0   \\  -3   &   -4   \end{pmatrix}$  
      & $\begin{pmatrix}  0   \\  -4                    \end{pmatrix}$  &
$23$  & $\begin{pmatrix}  0   &   3   \\  -8   &   -7   \end{pmatrix}$  
      & $\begin{pmatrix}  -10 \\  -4                    \end{pmatrix}$  &
$37$  & $\begin{pmatrix}  6   &   0   \\  -15  &   -6   \end{pmatrix}$  
      & $\begin{pmatrix}  -7  \\  -6                    \end{pmatrix}$ \\
 \myStrut
$10$  & $\begin{pmatrix}  3   &   1   \\  -7   &   -2   \end{pmatrix}$  
      & $\begin{pmatrix}  2   \\  4                     \end{pmatrix}$  &
$24$  & $\begin{pmatrix}  0   &   1   \\  -1   &   -1   \end{pmatrix}$  
      & $\begin{pmatrix}  3   \\  0                     \end{pmatrix}$  &
$38$  & $\begin{pmatrix}  0   &   1   \\  -1   &   -5   \end{pmatrix}$  
      & $\begin{pmatrix}  -6  \\  16                    \end{pmatrix}$  \\
 \myStrut
$11$  & $\begin{pmatrix}  1   &   1   \\  2    &   3    \end{pmatrix}$  
      & $\begin{pmatrix}  0   \\  5                     \end{pmatrix}$  &
$25$  & $\begin{pmatrix}  1   &   0   \\  6    &   1    \end{pmatrix}$  
      & $\begin{pmatrix}  3   \\  4                     \end{pmatrix}$  &
$39$  & $\begin{pmatrix}  7   &   2   \\  2    &   6    \end{pmatrix}$  
      & $\begin{pmatrix}  17  \\  14                    \end{pmatrix}$  \\
 \myStrut
$12$  & $\begin{pmatrix}  0   &   1   \\  -1   &   -3   \end{pmatrix}$  
      & $\begin{pmatrix}  3   \\  2                     \end{pmatrix}$  &
$26$  & $\begin{pmatrix}  9   &   0   \\  11   &   -23  \end{pmatrix}$  
      & $\begin{pmatrix}  2   \\  -7                    \end{pmatrix}$  &
$40$  & $\begin{pmatrix}  27  &   1   \\  14   &   -35  \end{pmatrix}$  
      & $\begin{pmatrix}  -19 \\  2                     \end{pmatrix}$  \\
 \myStrut
$13$  & $\begin{pmatrix}  4   &   2   \\  5    &   6    \end{pmatrix}$  
      & $\begin{pmatrix}  -6  \\  -5                    \end{pmatrix}$  &
$27$  & $\begin{pmatrix}  1   &   0   \\  10   &   -1   \end{pmatrix}$  
      & $\begin{pmatrix}  -4  \\  7                     \end{pmatrix}$  &
$41$  & $\begin{pmatrix}  18  &   0   \\  -5   &   16   \end{pmatrix}$  
      & $\begin{pmatrix}  1   \\  -15                   \end{pmatrix}$  \\
 \myStrut
$14$  & $\begin{pmatrix}  0   &   1   \\  -1   &   -1   \end{pmatrix}$  
      & $\begin{pmatrix}  -4  \\  3                     \end{pmatrix}$  &
$28$  & $\begin{pmatrix}  12  &   1   \\  -25  &   26   \end{pmatrix}$  
      & $\begin{pmatrix}  -6  \\  -8                    \end{pmatrix}$  &
$42$  & $\begin{pmatrix}  2   &   1   \\  11   &   -36  \end{pmatrix}$  
      & $\begin{pmatrix}  8   \\  7                     \end{pmatrix}$  \\
 \myStrut
$15$  & $\begin{pmatrix}  1   &   0   \\  5    &   -1   \end{pmatrix}$  
      & $\begin{pmatrix}  0   \\  7                     \end{pmatrix}$  &
$29$  & $\begin{pmatrix}  11  &   0   \\  -2   &   8    \end{pmatrix}$  
      & $\begin{pmatrix}  -4  \\  -2                    \end{pmatrix}$  &
$43$  & $\begin{pmatrix}  8   &   1   \\  -16  &   -18  \end{pmatrix}$  
      & $\begin{pmatrix}  14  \\  16                    \end{pmatrix}$  \\
\myStrut
$16$  & $\begin{pmatrix}  3   &   1   \\  -11  &   -14  \end{pmatrix}$  
      & $\begin{pmatrix}  5   \\  8                     \end{pmatrix}$  &
$30$  & $\begin{pmatrix}  10  &   1   \\  29   &   3    \end{pmatrix}$  
      & $\begin{pmatrix}  12  \\  1                     \end{pmatrix}$  &
$44$  & $\begin{pmatrix}  7   &   1   \\  -37  &   20   \end{pmatrix}$  
      & $\begin{pmatrix}  6   \\  19                    \end{pmatrix}$  \\
 \myStrut
$17$  & $\begin{pmatrix}  1   &   1   \\  2    &   3    \end{pmatrix}$  
      & $\begin{pmatrix}  8   \\  -4                    \end{pmatrix}$  &
$31$  & $\begin{pmatrix}  11  &   0   \\  6    &   -14  \end{pmatrix}$  
      & $\begin{pmatrix}  -5  \\  4                     \end{pmatrix}$  &
$45$  & $\begin{pmatrix}  1   &   0   \\  20   &   1    \end{pmatrix}$  
      & $\begin{pmatrix}  14  \\  -6                    \end{pmatrix}$  \\
 \myStrut 
$18$  & $\begin{pmatrix}  2   &   1   \\  7    &   -14  \end{pmatrix}$  
      & $\begin{pmatrix}  -3  \\  3                     \end{pmatrix}$  &
$32$  & $\begin{pmatrix}  27  &   1   \\  -8   &   -5   \end{pmatrix}$  
      & $\begin{pmatrix}  13  \\  -15                   \end{pmatrix}$ &
& & \\
\hline
\end{tabular}
\normalsize
\vspace{2 ex}
\caption{}
\label{tbl:Conjugacy}
\end{center}
\end{table}
 It is easily verified that
\begin{equation} 
[L_d,\boldsymbol{\eta}_d] [F_d,\boldsymbol{\chi}_d]
[L_d,\boldsymbol{\eta}_d]^{-1} 
= [Z,\boldsymbol{0}]
\end{equation} 
This means that if $\hat{U} \in
[L_d,\boldsymbol{\eta}_d]$, and if $|\psi_d\rangle$ is the RBSC vector in
dimension $d$, then $\hat{U}|\psi_d\rangle$ is a GP fiducial vector which
is an eigenvector of
$[Z,\boldsymbol{0}]$.  Conjecture B is thus confirmed numerically for
every dimension $\le 45$.

This suggests
\begin{quote}
\textbf{Conjecture C:}   GP fiducial vectors exist in every finite dimension. 
Furthermore, every such vector is an eigenvector of a canonical order $3$
unitary which is conjugate to
$[Z,\boldsymbol{0}]$. 
\end{quote}
Conjecture C is clearly stronger than Conjecture B.  It also implies Conjecture
A.

An operation conjugate to $[Z,\boldsymbol{0}]$ is automatically a canonical
order
$3$ unitary.   It would be interesting to know  whether the converse is also
true: 
\emph{i.e.}\ whether every  canonical order
$3$ unitary is conjugate to $[Z,\boldsymbol{0}]$.  If that
were  not the case  Conjecture C would be strictly stronger than Conjecture A.

\section{Dimensions $2$ to $7$:  Vectors, Orbits and Stability Groups}
\label{sec:VectorsOrbitsStability}
In dimensions $2$--$7$ RBSC made a numerical search, in an attempt to find the
total number of GP fiducial vectors.  On the assumption that their search
 was exhaustive 
 we use their data
to calculate, for dimensions 
$2$--$7$, the number of distinct orbits under the action of the extended Clifford
group.  We also calculate the order of the
 stability group corresponding to each orbit.  Our results are tabulated in
Table~\ref{tbl:Orbits}.  They confirm that in dimensions $2$--$7$
\emph{every} GP fiducial vector is an eigenvector of a canonical  order $3$
Clifford unitary (in agreement with Conjecture A).   We incidentally give
exact expressions for two of the GP fiducial vectors in dimension $7$ (one
on each of the two distinct orbits). 

The calculations on which these statements are based are somewhat lengthy, and
there is not the space to reproduce them here.  We therefore confine ourselves
to summarizing the  end results, which it is straightforward (albeit tedious) to
confirm with the help of (for example) 
\emph{Mathematica}.

\begin{table}
\begin{center}
\renewcommand{\arraystretch}{1.5}
\begin{tabular}{| c | c| c| c |}
\hline
& \multicolumn{2}{c |}{Stability Group} & \\
\cline{2-3}
dimension & type & order & number of orbits \\
\hline
$2$ & non-Abelian &  $6$ & $1$ \\
\hline
 & non-Abelian & $6$ & $\infty$ \\
\cline{2-4}
3 & non-Abelian & $12$ & $1$ \\
\cline{2-4}
 & non-Abelian & $48$ & $1$ \\
\hline
$4$ &  cyclic & $6$ & $1$ \\
\hline
$5$ & cyclic & $3$ & $1$ \\
\hline
$6$ & cyclic & $3$ & $1$ \\
\hline
& cyclic & $3$ & $1$ \\
\cline{2-4}
\raisebox{2 ex}[0 cm][0 cm]{$7$ } & cyclic & $6$ & $1$ \\
\hline
\end{tabular}
\end{center}
\vspace{ 0.1 in}
\caption{Stability groups in dimensions $2$--$7$.  In every case the
stability group includes an order
$3$ cyclic subgroup generated by a unitary operation having Clifford trace $=-1$.}
\label{tbl:Orbits}
\end{table}

\subsection*{Dimension 2}  Exact solutions in dimension  $2$ have been obtained
by Zauner~\cite{Zauner} and RBSC~\cite{Renes}. In dimension
$2$ the GP fiducial vectors all lie on a single orbit of the extended Clifford group. 
Consider the GP fiducial vector
\begin{equation}
|\psi_{2}\rangle = 
\sqrt{(3+\sqrt{3})/6} \; |e_0\rangle + e^{\frac{i
\pi}{4}}\sqrt{(3-\sqrt{3})/6} \; |e_1\rangle
\end{equation}
The stability group of $|\psi_2\rangle$ is the order~$6$, non-Abelian subgroup of
$\EC(2)/\Cc(2)$ generated by the unitary operation
\begin{equation}
[F_2,\boldsymbol{0}]
=
\left[
\begin{pmatrix} 0 & 1 \\ -1 & -1
\end{pmatrix},
\begin{pmatrix} 0 \\ 0
\end{pmatrix}
\right]
\end{equation}
and the three anti-unitary operations
{
\allowdisplaybreaks
\begin{align}
[A_2,\boldsymbol{0}]
& =
\left[
\begin{pmatrix} 0 & 1 \\ 1 & 0
\end{pmatrix},
\begin{pmatrix} 0 \\ 0
\end{pmatrix}
\right]
\\
[B_2,\boldsymbol{0}]
& =
\left[
\begin{pmatrix} -1 & -1 \\ 0 & 1
\end{pmatrix},
\begin{pmatrix} 0 \\ 0
\end{pmatrix}
\right]
\\
[C_2,\boldsymbol{0}]
& =
\left[
\begin{pmatrix} 1 & 0 \\ -1 & -1
\end{pmatrix},
\begin{pmatrix} 0 \\ 0
\end{pmatrix}
\right]
\end{align}
Note that $[F_2,\boldsymbol{0}]$ is
canonical  order $3$.  It follows from Lemmas~\ref{lem:CliffordStructure5}
and~\ref{lem:OrderECd} that $\left| \EC(2)/\Cc(2)\right| = 48$.  So the orbit
consists of $48 \div 6=8$ fiducial vectors (identifying vectors which only differ
by a phase), constituting $2$ distinct SIC-POVM's (as described by RBSC). 
}
\subsection*{Dimension 3} Exact solutions in dimension~$3$ have been obtained by
Zauner~\cite{Zauner} and RBSC~\cite{Renes}. We saw in
Section~\ref{sec:CliffordTrace} that dimension~$3$ is unusual in that it is the 
only dimension for which the identity operator has Clifford trace
$=-1$.   It seems to be unusual in another respect also:  for it is the only case
presently known where the GP fiducial vectors constitute infinitely many
distinct orbits of the extended Clifford group.

Consider the one parameter family of GP fiducial vectors
\begin{equation}
|\psi_{3}(t)\rangle = \frac{1}{\sqrt{2}} \left( e^{- i t} |e_1\rangle
-e^{i t} |e_2\rangle
\right)
\end{equation}
The complete set of GP fiducial vectors is obtained by acting on the vectors
$|\psi_{3}(t)\rangle$ with elements of $\EC(3)$.

Let $\hat{T}$ and $\hat{J}$ be the operators defined by Eqs.~(\ref{eq:TDef})
and~(\ref{eq:JopDef}) respectively.  Then
\begin{equation}
\hat{T} |\psi_{3}(t)\rangle =-|\psi_{3}({t+\tfrac{
\pi}{3}})\rangle
\qquad
\text{and}
\qquad
\hat{J} |\psi_{3}(t)\rangle = |\psi_{3}(-t)\rangle
\end{equation}
So $|\psi_{3}(t)\rangle$ and $|\psi_{3}(t')\rangle$ are on the same orbit if
$t'=\frac{ n \pi}{3} \pm t$ for some integer $n$.   At the cost of rather more
computational effort one can show that this condition is not only sufficient
 but also necessary for $|\psi_{3}(t)\rangle$ and $|\psi_{3}(t')\rangle$ to be on
the same  orbit.  So for each distinct orbit there is exactly one value of $t\in
[0,\frac{ \pi}{6}]$ such that $|\psi_{3}(t)\rangle$ is on the orbit.

There are three kinds of orbit:  a set of infinitely many generic orbits
corresponding to values of $t$ in the interior of the interval
$[0,\frac{\pi}{6}]$, and two exceptional orbits corresponding to the two end
points $t=0$ and $\frac{\pi}{6}$.

The stability group of the exceptional vector $|\psi_{3}(0)\rangle$ consists of
all
$48$ operations of the form $[F,\boldsymbol{0}]$, where $F$ is any element of
$\ESL(2,\mathbf{Z}_3)$.  The orbit thus consists of $432\div 48=9$ fiducial
vectors, constituting a single SIC-POVM.

The stability group of the exceptional vector $|\psi_{3}(\frac{\pi}{6})\rangle$ is
the order~$12$ non-Abelian subgroup of $\EC(3)/ \Cc(3)$ generated by the unitary
operation
\begin{equation}
[F_3,\boldsymbol{\chi}_3]=
\left[ \begin{pmatrix} -1 & 0 \\ -1 & -1
\end{pmatrix},
\begin{pmatrix} 0 \\ 1
\end{pmatrix}
\right]
\end{equation}
and the anti-unitary operation
\begin{equation}
[A_3,\boldsymbol{\chi}_3]=
\left[ \begin{pmatrix} 1 & 0 \\ 0 & -1
\end{pmatrix},
\begin{pmatrix} 0 \\ 1
\end{pmatrix}
\right]
\end{equation}
Note that
\begin{equation}
[F_3,\boldsymbol{\chi}_3]^2=
\left[ \begin{pmatrix} 1 & 0 \\ -1 & 1
\end{pmatrix},
\begin{pmatrix} 0 \\ 0
\end{pmatrix}
\right]
\end{equation} 
is canonical order $3$.  The orbit thus consists of 
$432\div 12=36$ fiducial
vectors, constituting $4$ distinct SIC-POVMs.

The stability group of a generic vector $|\psi_{3}(t)\rangle$ with
$0<t<\frac{\pi}{6}$ is the order~$6$ non-Abelian subgroup generated by the
unitary operation
\begin{equation}
[F_3,\boldsymbol{\chi}_3]^2=
\left[ \begin{pmatrix} 1 & 0 \\ -1 & 1
\end{pmatrix},
\begin{pmatrix} 0 \\ 0
\end{pmatrix}
\right]
\end{equation} 
 and the anti-unitary operation
\begin{equation}
[F_3,\boldsymbol{\chi}_3] \circ [A_3,\boldsymbol{\chi}_3]
=
\left[
\begin{pmatrix} -1 & 0 \\ -1 & 1
\end{pmatrix},
\begin{pmatrix} 0 \\ 0
\end{pmatrix}
\right]
\end{equation} 
The orbit thus consists of 
$432\div 6=72$ fiducial
vectors, constituting $8$ distinct SIC-POVMs.

\subsection*{Dimension 4}  The vector
\begin{multline}
|\psi_4\rangle  =
\sqrt{\frac{5-\sqrt{5}}{40}} \biggl(
2 \cos\frac{\pi}{8} |e_0\rangle
+ i \Bigl( e^{-\frac{i \pi}{8}} + \bigl(2+\sqrt{5}\bigr)^{\frac{1}{2}} 
e^{\frac{i
\pi}{8}}\Bigr) |e_1\rangle \biggr.
\\
\biggl.  
+ 2 i \sin \frac{\pi}{8} |e_2\rangle
+ i \Bigl( e^{-\frac{i \pi}{8}} - \bigl(2+\sqrt{5}\bigr)^{\frac{1}{2}} 
e^{\frac{i
\pi}{8}}\Bigr) |e_3\rangle
\biggr)
\end{multline}
is a GP fiducial vector in dimension~$4$, as discovered by Zauner~\cite{Zauner}
and RBSC~\cite{Renes}\footnote{
In Zauner's notation  $|\psi_4\rangle$ is the vector
\begin{equation}
e^{-\frac{i \pi}{8}} \left( X \psi_{1 a} + \rho^3 Y \psi_{1 b}\right)
\end{equation}
In  RBSC's notation  it is   the
vector 
\begin{equation}
r_0 |e_0\rangle + r_{+} e^{i \theta_{+}} |e_1\rangle + r_{1} e^{i
\theta_{1}} |e_2\rangle + r_{-} e^{i \theta_{-}} |e_3 \rangle
\end{equation}
for the case
$n= j=  m=1$ and $ k=0$ (note, however, that there is a typographical error in
RBSC~\cite{Renes}:  their expression for $r_0$ should read
$r_0 =
\sqrt{\left(1-1/\sqrt{5}\right)}/\left(2 \sqrt{2-\sqrt{2}}\right)$). 
}.
The stability group of $|\psi_4\rangle$ is the order~$6$ cyclic subgroup of
$\EC(4)/\Cc(4)$ generated by the anti-unitary operation
\begin{equation}
[A_4,\boldsymbol{\chi}_4]
=
\left[ 
\begin{pmatrix} -1 & 1 \\ -1 & 2
\end{pmatrix},
\begin{pmatrix} 2 \\ 0
\end{pmatrix}
\right]
\end{equation}
Note that 
\begin{equation}
[A_4,\boldsymbol{\chi}_4] ^2
=
\left[ 
\begin{pmatrix} 0 & 1 \\ -1 & 3
\end{pmatrix},
\begin{pmatrix} 0 \\ 2
\end{pmatrix}
\right]
=
\left[ 
\begin{pmatrix} 0 & 1 \\ -1 & -1
\end{pmatrix},
\begin{pmatrix} 0 \\ 0
\end{pmatrix}
\right]
\end{equation}
is canonical order $3$  (where we used
Eq.~(\ref{eq:KernelExpression}) to obtain the last expression on the right hand
side).

It follows from Lemmas~\ref{lem:CliffordStructure5} and~\ref{lem:OrderECd} that
the group $\EC(4)/\Cc(4)$ is of order $1536$.  So the orbit generated by 
$|\psi_4\rangle$ contains $1536 \div 6= 256$ fiducial vectors, constituting $256
\div 16=16$  SIC-POVMs.   It was shown by RBSC that there are only $16$
 SIC-POVMs in dimension $4$.  We conclude that the fiducial vectors all
lie on a single orbit of the extended Clifford group. 

\subsection*{Dimension 5}  Let $|\psi_5\rangle$ be RBSC's numerical vector in
dimension~$5$.  We noted in the last section that the stability group of
$|\psi_5\rangle$ is of order $3$.  It follows from 
 Lemmas~\ref{lem:CliffordStructure5} and~\ref{lem:OrderECd} that
the group $\EC(5)/\Cc(5)$ is of order $6000$.  So the orbit generated by 
$|\psi_5\rangle$ contains $6000\div 3 = 2000$ fiducial vectors, constituting
$2000\div 25=80$ SIC-POVMs.  It was shown by RBSC that there are only $80$
SIC-POVMs in dimension~$5$. We conclude that the fiducial vectors all lie on a
single orbit of the extended Clifford group.

Note that  Zauner's analytic solution in dimension $5$ (on p.~63 of his
thesis~\cite{Zauner}) can be used to give exact expressions for each of the
vectors on the orbit.

\subsection*{Dimension 6}  Let $|\psi_6\rangle$ be RBSC's numerical vector in
dimension~$6$.  We noted in the last section that the stability group of
$|\psi_6\rangle$ is of order $3$.  It follows from 
 Lemmas~\ref{lem:CliffordStructure5} and~\ref{lem:OrderECd} that
the group $\EC(6)/\Cc(6)$ is of order $10368$.  So the orbit generated by 
$|\psi_6\rangle$ contains $10368\div 3 = 3456$ fiducial vectors, constituting
$3456\div 36=96$ SIC-POVMs.  It was shown by RBSC that there are only $96$
SIC-POVMs in dimension~$5$. We conclude that the fiducial vectors all lie on a
single orbit of the extended Clifford group (in agreement with 
Grassl's~\cite{Grassl}  analysis, based on his exact solution in dimension $6$).

Note that Grassl's~\cite{Grassl} analytic solution can be used to give exact
expressions for each of the vectors on the orbit.

\subsection*{Dimension 7}  Let $|\psi_7\rangle$ be RBSC's numerical vector in
dimension~$7$.  We noted in the last section that the stability group of
$|\psi_7\rangle$ is of order $6$.  It follows from 
 Lemmas~\ref{lem:CliffordStructure5} and~\ref{lem:OrderECd} that
the group $\EC(6)/\Cc(6)$ is of order $32928$.  So the orbit generated by 
$|\psi_7\rangle$ contains $32928\div 6 = 5488$ fiducial vectors, constituting
$5488\div 49=112$ SIC-POVMs.  However, it was shown by RBSC that there are $336$
SIC-POVMs in dimension $7$.  We conclude that there must be at least one other
orbit.

The search for the additional orbit or orbits is 
facilitated by the fact that in dimension
$7$ there exists a canonical order $3$ Clifford unitary for which the $F$ matrix
is diagonal:  namely
\begin{equation}
[F'_{7},\boldsymbol{0}]
=
\left[\begin{pmatrix} - 3 & 0 \\ 0 & 2
\end{pmatrix},
\begin{pmatrix}  0 \\ 0
\end{pmatrix}
\right]
\end{equation}
The fact that $F'_{7}$ is diagonal means that the $\hat{U}\in
[F'_{7},\boldsymbol{0}]$ are permutation matrices.  Specifically
\begin{equation}
\hat{U} = e^{i \theta} \sum_{r=0}^{6} |e_{4 r}\rangle \langle e_r |
\end{equation}
for every $\hat{U}\in
[F'_{7},\boldsymbol{0}]$
(where $e^{i \theta}$ is an arbitrary phase, and where we have used the
decomposition described in Lemma~\ref{lem:CliffordStructure3}).  This
considerably simplifies the calculations.
 We will also have occasion to consider the anti-unitary operation
\begin{equation}
[A'_{7},\boldsymbol{0}]
=
\left[\begin{pmatrix} -2 & 0 \\ 0 & -3
\end{pmatrix},
\begin{pmatrix}  0 \\ 0
\end{pmatrix}
\right]
\end{equation}
which is a square root of $[F'_{7},\boldsymbol{0}]$.

We look for eigenvectors of $[F'_{7},\boldsymbol{0}]$.
Let
\begin{equation}
l_r = \begin{cases} 
1 \qquad & \text{if $r=1,2$ or $4$} \\
-1 & \text{if $r=3,5$ or $6$}
\end{cases}
\end{equation}
Also let
\begin{equation}
 a_0 = \frac{1}{2} \left(\sqrt{\frac{1}{4-\sqrt{2}}} +
i \sqrt{\frac{4-\sqrt{2}}{2}}
\right) \qquad a_1 = \frac{1}{4} \sqrt{\frac{8-5\sqrt{2}}{7}}
\qquad a_2 = 2^{-\frac{7}{4}}
\end{equation}
and
\begin{equation}
b_0 = \sqrt{\frac{2+3\sqrt{2}}{14}} \qquad
b_1=\sqrt{\frac{4-\sqrt{2}}{28}} \qquad
\theta = \cos^{-1} \left(-\frac{\sqrt{\sqrt{2}+1}}{2} \right)
\end{equation}
Then define
\begin{align}
|\psi'_7\rangle & = a_0 |e_0\rangle -\sum_{r=1}^{6} \left(a_1 + l_r a_2\right)
|e_r\rangle 
\label{eq:Dim7AnalyticVectorA}
\\
|\psi''_7\rangle & = b_0 |e_0\rangle + \sum_{r=1}^{6} b_1 e^{i l_r \theta}
|e_r\rangle
\label{eq:Dim7AnalyticVectorB}
\end{align}
It is readily confirmed that $|\psi'_7\rangle$ and $|\psi''_7\rangle$ are both GP
fiducial vectors.  The stability group of $|\psi'_7\rangle$ is the order $3$
subgroup generated by $[F'_{7},\boldsymbol{0}]$, while the stability group of 
$|\psi''_7\rangle$ is the order $6$ subgroup generated by
$[A'_{7},\boldsymbol{0}]$.  Since the stability groups are non-isomorphic the
orbits generated by $|\psi'_7\rangle$ and $|\psi''_7\rangle$ are disjoint.  The
orbit generated by 
$|\psi'_7\rangle$ contains $32928\div 3 = 10976$ fiducial vectors, constituting
$10976\div 49=224$ SIC-POVMs.  The
orbit generated by 
$|\psi''_7\rangle$ contains  $5488$ fiducial vectors, constituting a further
$112$ SIC-POVMs.  This accounts for all $336$  of the SIC-POVMs identified by
RBSC.  We conclude that there are no other orbits, apart from these two.

For the sake of completeness let us note that
\begin{equation}
|\psi_7\rangle = \hat{U} |\psi''_7\rangle
\end{equation}
where $|\psi_7\rangle$ is RBSC's numerical vector and $\hat{U}$ is a unitary
operator
\begin{equation}
\hat{U} \in 
\left[ \begin{pmatrix} 1 & 1 \\ -3 & -2
\end{pmatrix},
\begin{pmatrix} 0 \\1
\end{pmatrix}
\right]
\end{equation}

Finally, let us remark that $l_r$ is the Legendre symbol (see, \emph{e.g.},
Nathanson~\cite{Nathanson} or Rose~\cite{Rose})
\begin{equation}
l_r=\genfrac{(}{)}{}{}{r}{7}
\end{equation}
It has the important property that $l_{rs}=l_r l_s$ for all $r,s\in \mathbb{Z}$.

\section{A Fiducial Vector in Dimension 19}
\label{sec:dimension19}
In Section~\ref{sec:RBSCVectors} we saw that, except in dimension~$7$, each of
RBSC's numerical solutions has stability group of order~$3$.  This might
encourage one  to speculate  that when
$d>7$ the stability group is \emph{always} of order~$3$.  In this section we
 show that there is at least one exception to that
putative rule, by constructing a GP fiducial vector in dimension~$19$ for which
the stability group has order $\ge 18$.

The vector we construct is an eigenvector of the order $18$ anti-unitary operation
\begin{equation}
[A'_{19},\boldsymbol{0}]
=
\left[
\begin{pmatrix} -9 & 0 \\ 0 & -2
\end{pmatrix},
\begin{pmatrix} 0\\ 0
\end{pmatrix}
\right]
\in
\EC(19)/\Cc(19)
\end{equation}
Note that
\begin{equation}
[F'_{19},\boldsymbol{0}]
=
[A'_{19},\boldsymbol{0}]^6
=\left[
\begin{pmatrix} -8 & 0 \\ 0 & 7
\end{pmatrix},
\begin{pmatrix} 0\\ 0
\end{pmatrix}
\right]
\end{equation}
is canonical order~$3$.

The construction is similar  to  our
construction of the vector
$|\psi''_7\rangle$ in the last section.  Let
$l'_r$ be the Legendre symbol
\begin{equation}
l'_r = \genfrac{(}{)}{}{}{r}{19}
= \begin{cases} 1 \qquad & \text{if $r=1,4,5,6,7,9,11,16$ or $17$} \\
-1\qquad & \text{if $r=2,3,8,10,12,13,14,15$ or $18$}
\end{cases} 
\end{equation}
and let
\begin{equation}
b'_0 = \sqrt{\frac{5 + 9\sqrt{5}}{95} } \qquad
b'_1= \sqrt{\frac{10-\sqrt{5}}{190}} \qquad
\theta'= \cos^{-1} \left(\sqrt{\frac{\sqrt{5}-1}{8}} \right)
\end{equation}
Then define
\begin{equation}
|\psi'_{19}\rangle = b'_0 |e_0\rangle
+ \sum_{r=1}^{18} b'_1 e^{ i l'_r \theta'} |e_r\rangle
\label{eq:Dim19AnalyticVector}
\end{equation}
It is readily confirmed that $|\psi'_{19}\rangle$ is a GP fiducial vector, and an
eigenvector of $[A'_{19},\boldsymbol{0}]$.  

Observe that the orbit generated by $|\psi'_{19}\rangle$ is disjoint from the
orbit generated by RBSC's numerical vector  $|\psi_{19}\rangle$ (because the
stability groups are non-isomorphic).  It follows that there are at least two
distinct orbits in dimension
$19$.

\section{Diagonalizing the $F$ matrix}
\label{sec:diagonalF}
Our construction of the exact solutions $|\psi'_7\rangle$,  $ | \psi''_7\rangle$
and 
$|\psi'_{19}\rangle$ in Eqs.~(\ref{eq:Dim7AnalyticVectorA}), 
(\ref{eq:Dim7AnalyticVectorB}) and~(\ref{eq:Dim19AnalyticVector}) was facilitated
by the fact that in dimensions $7$ and $19$ there exist canonical order $3$
unitaries for which the corresponding $F$ matrix is diagonal.    It is natural to
ask in what other dimensions that is true.  The  theorem proved below
 answers that question.

We will need the following lemma:
\begin{lemma}
\label{lem:DiagonalF}
Let $p$ be a prime number $=1\;(\text{mod}\; 3) $, and let $n$ be any 
integer $\ge 1$.  Then there exists an integer $\alpha$ such that
\begin{equation}
 \alpha^2 + \alpha + 1 =  0 \quad (\text{mod}\; p^n)
\end{equation}
\end{lemma}
\begin{proof}
The proof relies heavily on the theory of primitive roots, as described in (for
example) Chapter 3 of Nathanson~\cite{Nathanson} or Chapter~5 of
Rose~\cite{Rose}.  Let $\phi$ be Euler's phi, or totient function (so for every
 integer $x\ge 1$, $\phi(x)$ is the number of  integers $y$ in the range $1\le y<
x$ which are relatively prime to $x$).  Then there exists a single positive
integer
$g$ such that for every integer $m\ge 1$ the multiplicative order of $g$,
considered as an element of $\mathbb{Z}_{p^m}$, is $\phi(p^m)=(p-1)p^{m-1}$ (see,
for example, Nathanson~\cite{Nathanson}, p.~93, or Rose~\cite{Rose}, p.~91).  The
fact that $p=1\;(\text{mod}\; 3)$ means $p= 3 k + 1$        for some  integer
$k\ge 1$.  Define
\begin{equation}
\alpha = g^{k p^{n-1} }
\end{equation}
It is then immediate that 
\begin{equation}
\alpha^3 = g^{\phi(p^n)} = 1 \quad (\text{mod}\; p^n)
\label{eq:Lem8AlphaCubed}
\end{equation}

It is also true that $\alpha-1$ is relatively prime to $p$.  For suppose that
were not the case.   It would then follow from the definition of $\alpha$, and
the fact that $g$ is a primitve root \emph{modulo} $p$, that 
\begin{equation}
  k p^{n-1} = l (p-1) = 3 k l
\end{equation}
for some integer $l\ge 1$.  That, however, is impossible since $p$ is not a
multiple of $3$.

The fact that $\alpha - 1$ is relatively prime to $p$ means that there exists an
integer $\beta$ such that
\begin{equation}
 \beta (\alpha - 1) = 1 \quad (\text{mod}\; p^n)
\label{eq:Lem8AlphaMinusOne}
\end{equation}
It now follows from Eqs.~(\ref{eq:Lem8AlphaCubed})
and~(\ref{eq:Lem8AlphaMinusOne}) that 
\begin{equation}
\alpha^2 + \alpha + 1 = \beta (\alpha^3 -1 ) = 0 \quad (\text{mod}\; p^n)
\end{equation}
\end{proof}

We are now in a position to prove our main result:
\begin{theorem}
There exists a canonical order $3$ unitary $[F,\boldsymbol{\chi}] \in
\C(d)/\Cc(d)$ for which the matrix $F$ is diagonal if and only if each of the
following is true
\begin{enumerate}
\item $d$ has at least one prime divisor $=1\; (\text{mod} \; 3)$.
\item  $d$ has no prime divisors $=2\; (\text{mod}\; 3)$.
\item  $d$ is not divisible by $9$.
\end{enumerate} 
\end{theorem}
\begin{remark}
So there exist canonical order $3$ unitaries $[F,\boldsymbol{\chi}]$ for which
$F$ is diagonal in dimension $7, 13, 19, 21, 31, 37, 39, 43, 49,  \dots$
\end{remark}
\begin{proof}
We begin by proving sufficiency.  Suppose that conditions (1), (2) and (3) are all
true.  Then we have, for some $t\ge 1$,
\begin{equation}
d= 3_{\vphantom{1}}^{n_0} p_1^{n_1} \dots p_t^{n_t}
\end{equation}  
where the $p_i$ are distinct prime numbers $=
1\;(\text{mod} 3)
$, where the integer $n_0 =0$ or $1$, and where the integers $n_1, \dots ,
n_t$ are all
$\ge 1$.  It follows from Lemma~\ref{lem:DiagonalF} that there exist integers
$\alpha_1, \dots , \alpha_t$ such that 
\begin{equation}
\alpha_i^2 + \alpha^{\vphantom{2}}_i + 1 = 0 \quad (\text{mod}\; p_i^{n_i})
\end{equation}
for $i=1, \dots, t$.  We  then use the Chinese remainder theorem (see, for
example, Nathanson~\cite{Nathanson} or Rose~\cite{Rose}) to deduce that there
exists a single integer $\alpha$ such that
\begin{align}
 \alpha & = 1 \quad (\text{mod}\; 3)
\\
\intertext{and}
\alpha & = \alpha_i \quad (\text{mod}\; p_i^{n_i})
\end{align}
for $i=1, \dots , t$.  We have
\begin{align}
 \alpha^2 + \alpha + 1 & = 0 \quad (\text{mod}\; 3)
\\
\intertext{and}
\alpha^2 + \alpha + 1 & = 0  \quad (\text{mod}\; p_i^{n_i})
\end{align}
for $i=1, \dots , t$.  Consequently
\begin{equation}
\alpha^2 + \alpha + 1 = 0 \quad (\text{mod}\; d)
\end{equation}
It follows that the matrix
\begin{equation}
F = \begin{pmatrix} \alpha & 0 \\ 0 & -\alpha - 1
\end{pmatrix}
\end{equation}
$\in \SL(2, \mathbb{Z}_{\overline{d}})$
(bearing in mind that $d$ is odd).  Moreover, $\Tr(F) = -1 \; (\text{mod} \; d)$.
Since $d \neq 3$ we conclude that $[F, \boldsymbol{\chi}]$ is a canonical order
$3$ unitary for all $\boldsymbol{\chi} \in (\mathbb{Z}_d)^2$.  This proves
sufficiency.

To prove necessity suppose
\begin{equation}
F 
=
\begin{pmatrix} \alpha & 0 \\ 0 & \delta
\end{pmatrix}
\in
\SL(2,\mathbb{Z}_{\overline{d}} )
\end{equation}
is such that $[F, \boldsymbol{\chi}]$ is canonical order $3$ for some
$\boldsymbol{\chi} \in (\mathbb{Z}_d)^2$.  Then  $\alpha +
\delta = -1\; (\text{mod} \; d)$, implying  
\begin{align}
\alpha^2 + \alpha + 1 & = 0 \quad (\text{mod} \; d)
\label{eq:alphaQuadraticCondition}
\\
\alpha^3 & = 1 \quad (\text{mod} \; d)
\label{eq:alphaCubedCondition}
\end{align}
(in view of the fact that $\alpha \delta = 1 \; (\text{mod} \; \overline{d})$).

To show that $d$ has no prime divisors $=2\; (\text{mod}\; 3)$ assume  the
contrary. It  would
then follow from Eqs.~(\ref{eq:alphaQuadraticCondition})
and~(\ref{eq:alphaCubedCondition}) that
\begin{align}
\alpha^2 + \alpha + 1 & = 0 \quad (\text{mod} \; p)
\label{eq:alphaQuadraticConditionB}
\\
\alpha^3 & = 1 \quad (\text{mod} \; p)
\label{eq:alphaCubedConditionB}
\end{align}
for some prime number  $p=2\; (\text{mod}\; 3)$.  Let $r$ be a
primitive root of
$p$  and let
$k\in \mathbb{Z}$  be such that
$0\le k < p-1$ and $\alpha = r^k\;(\text{mod} \; p)$ (see, for example,
Nathanson~\cite{Nathanson} or Rose~\cite{Rose}).  Then
Eq.~(\ref{eq:alphaCubedConditionB}) implies
$r^{3 k}=1\;(\text{mod} \; p)$ which, in view of the fact that $r$ is a primitive
root, means $3 k = l( p-1)$ for some  $l\in \mathbb{Z}$.  The fact that $0\le k <
p-1$ implies $ 0 \le l < 3 $.  Taking into account the fact that $p-1$ is not
divisible by $3$ (because $p=2\; (\text{mod}\; 3)$)  we deduce that $l=0$. 
But then $k= 0$, implying $\alpha=1 \;(\text{mod} \; p)$.  In view of
Eq.~(\ref{eq:alphaQuadraticConditionB}) this means $3=0 \;(\text{mod}\; p)$: 
which is a contradiction.

To prove that $d$ is not divisible by $9$ we again proceed by \emph{reductio ad
absurdum}.   Suppose that $d$ were divisible by $9$.  It would then follow
from Eq.~(\ref{eq:alphaQuadraticCondition}) that
\begin{equation}
\alpha^2 + \alpha + 1 = 0 \quad (\text{mod} \; 9)
\label{eq:alphaQuadraticConditionC}
\end{equation}
However, it is easily verified (by explicit enumeration) that this equation has
no solutions.

Finally, suppose that $d$ had no prime divisors $=1 \;(\text{mod}\; 3)$.  In view
of the results just proved it would  follow that $d=3$.  But if $d=3$,
Eq.~(\ref{eq:alphaQuadraticCondition}) implies $\alpha = 1\; (\text{mod} \; 3)$. 
Taking into account the requirement  $\alpha \delta =\det F = 1 \; (\text{mod} \;
3)$ this means
$\delta = 1 \; (\text{mod} \; 3)$.  But then $F$ is the identity matrix,
which contradicts  the assumption that $[F,\boldsymbol{\chi}]$ is a canonical
order
$3$ unitary.  We conclude that
$d$ must have at least one prime divisor $=1 \;(\text{mod}\; 3)$. 
\end{proof}
\section{Conclusion}
RBSC conclude their paper by saying ``a rigorous proof of existence of SIC-POVMs
in all finite dimensions seems tantalizingly close, yet remains somehow
distant''.  That well expresses our own perception of the matter.  While working
on this problem we have several times had the sense that the crucial discovery
lay just round the corner, only to find that our hopes were illusory.  We make
our results public in the hope that they may, nevertheless, contain a few clues,
which will help to take us  further forward.

In particular it seems to us that significant progress would be made if it could
be established whether it is in fact true that every GP fiducial vector is an
eigenvector of a canonical order $3$ unitary.  Also, if that is the case, one
would like to know exactly
\emph{why} it is the case. 
\subsubsection*{Acknowledgements}  I am grateful to Chris Fuchs for exciting my
interest in this problem.  I am also grateful to him and to Paul Busch, Barry
Rhule and R\"{u}diger Schack for numerous inspiring discussions about POVMs in
particular, and the mysteries of quantum mechanics in general.  Finally, I am
grateful to Markus Grassl, Gerhard Zauner and  Alexander Vlasov for some
very helpful comments regarding the first version of this paper.

\end{document}